\documentclass[%
 reprint,
 amsmath,amssymb,
 aps,
 pra,
 longbibliography,
 floatfix,
]{revtex4-2}

\usepackage{graphicx}
\usepackage{dcolumn}
\usepackage{bm}
\usepackage{booktabs}
\usepackage{tcolorbox}
\usepackage[colorlinks=true, allcolors=blue]{hyperref}

\AtBeginDocument{\hbadness=10000\hfuzz=100pt}

\begin{document}
\nocite{apsrev42Control}

\title{LLM-as-a-judge validity is strongly task-dependent across physics assessment formats}

\author{Will Yeadon}
\affiliation{Department of Physics, Durham University, Durham DH1 3LE, UK}

\author{Tom Hardy}
\affiliation{Department of Physics, Durham University, Durham DH1 3LE, UK}

\author{Paul Mackay}
\affiliation{Department of Physics, Durham University, Durham DH1 3LE, UK}

\author{Elise Agra}
\affiliation{Department of Physics, Durham University, Durham DH1 3LE, UK}

\date{\today}

\begin{abstract}
As large language models (LLMs) are increasingly considered for automated assessment and feedback, understanding when LLM marking is valid is essential. We evaluate LLM-as-a-judge marking across four settings spanning three assessment formats - structured questions, written essays, and scientific plots - comparing GPT-5.2, Grok 4.1, Claude Opus 4.5, DeepSeek-V3.2, Gemini 3 Pro, and committee aggregations against human markers under blind, solution-provided, false-solution, and anchored conditions. We distinguish absolute accuracy from rank-order agreement, since a marking system can match the distribution of human marks while failing to order responses by quality. Across task types, performance is sharply task-dependent. For blind university exam questions ($n=771$) and secondary and university structured questions ($n=1151$), models show robust rank-order agreement with human markers (Spearman $\rho > 0.6$), with official solutions reducing error and strengthening agreement. False solutions degrade absolute accuracy, showing that models defer to provided references, but leave rank-ordering intact. Essay marking behaves fundamentally differently. Across $n=55$ scripts, each containing five essays ($n=275$ essays total), blind AI marking is harsher and more variable than human marking and adding marking guidance does not improve rank-order agreement. Anchored exemplars shift the AI mean close to the human mean and compress variance below the human standard deviation, but rank-order agreement remains near-zero. For code-based plot elements ($n=1400$), models achieve high rank-order agreement ($\rho > 0.83$) with near-linear calibration. LLM marking validity depended strongly on the assessment task, and this held for every contemporary model tested. The results also show that the reliability of the human benchmark constrains the claims that can be made about AI-human agreement.
\end{abstract}

\maketitle

\section{\label{sec:intro}Introduction}

As large language models (LLMs) become increasingly capable of solving physics problems, attention naturally turns to whether these models can also validly assess student work. Early evaluations showed GPT-3.5 could not reliably pass an introductory physics course~\cite{kortemeyer2023could_ai_pass}, but recent models now surpass human participants on structured Physics Olympiad problems~\cite{tschisgale2025olympiad}, and multimodal systems outperform average post-instruction undergraduates on standard concept inventories~\cite{kortemeyer2025multilingual}. Thus, if generating correct answers is now straightforward for modern LLMs, the question becomes whether LLMs can also be used for marking and feedback. Using LLMs to evaluate answers is typically referred to as ``LLM-as-a-judge''~\cite{laaj-survey}. This work considers whether LLM-as-a-judge systems could be used in physics education as a fast and scalable assessment tool.

There is a growing push to use AI for improved feedback~\cite{thakkar2026large}, but the practical applicability is strictly constrained by governance. In England, Ofqual has explicitly stressed that using AI as the sole mechanism for awarding marks is not compliant with current regulations~\cite{ofqual2024ai_principles_marking}. Similarly, the European Union's AI Act treats AI systems intended to evaluate learning outcomes as a ``high-risk'' use category~\cite{eu2024_ai_act}. If AI is to be used as an assistant or auditor, the core danger is not merely average marking error, but undetected systematic bias that would undermine fairness and trust.

Early work on LLM grading of physics problems has demonstrated feasibility at the level of introductory courses~\cite{kortemeyer2023toward}, and subsequent psychometric analysis has investigated when agreement with human markers can be trusted~\cite{kortemeyer2024psychometrics}. Subsequent work on a real 252-student thermodynamics exam found that fine-grained rubrics can paradoxically reduce AI grading accuracy~\cite{kortemeyer2024handwritten_thermo}, while comparisons of conventional machine learning and LLMs for classifying student concept use in physics show that LLMs offer competitive performance with far less task-specific engineering~\cite{kieser2024david_goliath}. Partial-credit grading of physics explanations has shown that LLMs can approach human-level accuracy on structured items, but that confidence varies substantially across question types~\cite{chen2024partial_credit_grading_confidence}. In higher education more broadly, comparisons between human and LLM grading of written exams find that aggregate-level agreement can mask significant item-level discrepancies~\cite{floden2025grading_exams_chatgpt}, while validity and reliability analyses of automated essay scoring reveal that apparent distributional agreement with human marks does not guarantee stable or valid scoring~\cite{pack2024llm_aes_validity_reliability}. These findings converge on a concern already well-established in the general LLM-as-a-judge literature: aggregate metrics such as mean absolute error or correlation can be misleading because LLM judges may exhibit systematic prompt-conditioned biases, including anchoring to supplied references~\cite{zheng2023mtbench_chatbot_arena_llm_judge,laaj-survey,lou2024anchoring}.

A critical but underexplored question in the physics education context is whether these failure modes are task-specific or general. Physics degree programmes require student output across qualitatively different modalities - structured mathematical derivations, scientific writing, and computational or visual artefacts - and it is plausible that AI marking reliability varies substantially across these formats. Reference materials and exemplars can dramatically reshape AI marking behaviour, and a model may become an anchored distributor (distributing marks around a specific numerical anchor) when faced with open-ended prose. Further, this anchoring could potentially be influenced by showing the LLM a range of different examples of work receiving specific marks. When adding marking support converts an LLM from a genuine evaluator into a distribution-matcher, low mean absolute error (MAE) is not only insufficient as a validity criterion~\cite{messick1995validity} but can be actively misleading~\cite{shankar2024validate_the_validators}. More broadly, recent work has emphasised that a correct final answer should not be conflated with the quality of the reasoning used to reach it~\cite{yeadon2026answer}.

In this work, we evaluate LLM-as-a-judge marking across four settings spanning three assessment formats drawn from a single undergraduate physics programme, each representing a major category of student output. This design reduces variation in disciplinary context while allowing assessment formats to be compared within physics. This work contributes to a growing body of PER scholarship applying NLP and machine-learning methods to physics assessment and qualitative analysis~\cite{wilson2022nlp_classification, tschisgale2023cgt, odden2024embeddings, fussell2024trustworthy_claims}. 

We address two research questions:
\begin{enumerate}
    \item How do LLM marking errors and rank-order agreement with human markers compare across structured questions, essays, and scientific plots? By rank-order agreement we mean the degree to which the marking
    system places responses in the same quality order as human markers - the property required whenever marks are used to rank, grade, or determine progression.
    \item Within each task type, how does the informational structure of the marking prompt - the presence or absence of a model solution, a false solution, or anchored exemplars - alter both absolute accuracy and rank-order agreement?
\end{enumerate}

The first setting comprises 771 exam questions from Durham University (2018--2022) across ten modules ranging from foundational physics to advanced condensed matter and theoretical astrophysics. Under institutional constraints, solutions are not distributed alongside this data, forcing any automated system to operate blind. This mirrors the likely deployment setting in many departments where question security prevents releasing schemes, making ``blind'' automated marking the realistic baseline rather than an artificial handicap. Prior work in physics education research also shows that human graders often evaluate solutions based on visible reasoning and domain knowledge rather than strict rubric matching, especially across different instructional contexts~\cite{marshman2017contrasting_grading_approaches}.

The second setting uses datasets where an answer key is available: OCR GCSE and A-Level Physics (2017--2021) and standard university textbook questions. These datasets allow us to test both blind and solution-informed marking, quantifying the marginal value of giving models access to official reference solutions. We also include false solutions to isolate whether models independently verify the physics or simply match text patterns against the provided standard. For the structured-question datasets, the marked responses are AI-generated rather than authentic student submissions, so these results should be interpreted as evidence about marking validity on a controlled response pool, not as a direct estimate of classroom deployment performance.

\begin{figure}[h]
\centering
\includegraphics[width=\columnwidth]{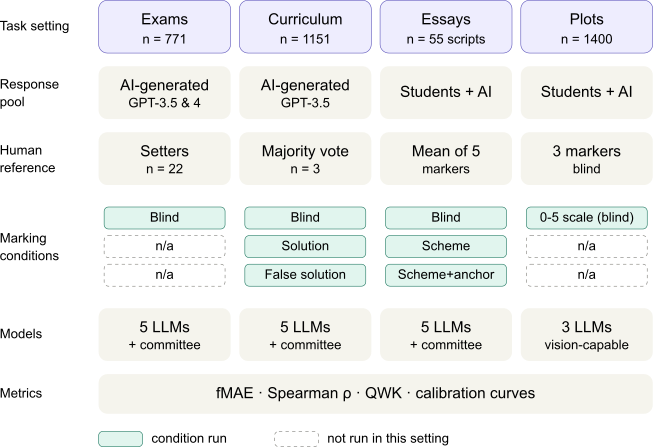}
\caption{\label{fig:study-design}Study design overview showing the four task settings and marking conditions used. fMAE denotes fractional mean absolute error and QWK quadratic weighted kappa.}
\end{figure}

The third and fourth settings comprise $n = 275$ physics essays combined into 55 scripts and $n = 1400$ individual scientific plots, representing the written and computational output modalities of undergraduate physics. The four settings and conditions are shown in FIG.~\ref{fig:study-design}. All settings use digitally generated artefacts rather than handwritten text, handwritten working, or hand-drawn diagrams. The conclusions of this study are therefore conditional on student work being available either natively in digital form or as a losslessly converted digital artefact.

\section{\label{sec:theory}Theoretical Background}
\subsection{Physics assessment task types}
In a modern Physics degree programme there are many diverse assessment types used to evaluate student understanding, including invigilated examinations, take-home assignments, laboratory reports, essays, viva examinations, and computational notebook projects. These formats vary significantly in structural constraints ranging from mathematical derivations with a unique target solution to open-ended writing requiring holistic qualitative evaluation \cite{docktor2016problem_solving_rubric}. Psychometric investigations into physics grading cultures reveal that human inter-rater reliability is strongly dependent on this task format \cite{marshman2017contrasting_grading_approaches, kortemeyer2024psychometrics}.

\subsection{Automated scoring}
Automated Short-Answer Grading (ASAG) in science and mathematics historically relied on rule-based NLP systems and supervised machine learning pipelines featuring manual feature extraction \cite{gonzalez2026interpretable, wilson2022nlp_classification}. Within physics education research, these classical machine learning classifiers and text embeddings have been extensively applied to classify student concept use and analyse qualitative data at scale \cite{kieser2024david_goliath, odden2024embeddings}. Whilst more interpretable than LLMs, traditional machine learning implementations require large task-specific labelled datasets for training and generally fail to generalize to novel open-response prompts or complex multi-step physics derivations without extensive engineering \cite{gonzalez2026interpretable, wilson2022nlp_classification}. This task-specific engineering burden is precisely the cost that LLM-as-a-judge evaluation promises to remove.

\subsection{LLM-as-a-judge}
The deployment of large language models (LLMs) as evaluators offers a scalable alternative to traditional automated scoring pipelines \cite{laaj-survey, walsh2026combining}. Recent evaluations confirm that contemporary models can automate the grading of complex physics calculations using advanced prompt structures such as Chain-of-Thought \cite{wei2025research}. In physics education, providing fine-grained rubrics or solutions can paradoxically distort grading accuracy, as models frequently defer to provided text patterns rather than independently verifying the underlying physical logic \cite{kortemeyer2024handwritten_thermo, lou2024anchoring}. Furthermore, the efficacy of LLM assessment varies drastically by response modality. While LLMs excel at categorical classification and structured text evaluation, traditional supervised machine learning pipelines or human consensus panels frequently outperform zero-shot LLM prompts when assessing continuous, holistic writing quality, such as in academic essays \cite{kim2025chatgpt, pack2024llm_aes_validity_reliability}. This difference necessitates evaluating automated marking systems on both absolute error profiles and validity across distinct structural tasks \cite{messick1995validity, shankar2024validate_the_validators}. Because model capabilities shift rapidly between releases, findings tied to a particular model generation may not transfer directly to later ones. We therefore evaluate several contemporary models and report cross-model as well as committee performance rather than treating any single model as representative of LLM-as-a-judge behaviour.

\subsection{\label{sec:theory-dist-vs-rank}Validity, reliability, and rank-order agreement}
Reliability and validity are related but distinct. \emph{Reliability} concerns whether the same work receives the same mark across markers or repeated gradings. In this study, \emph{validity} is evaluated in the bounded sense of evidence that LLM-generated marks agree with the human assessment reference, both in absolute score and in rank ordering. These analyses therefore provide evidence relative to the human benchmark; they do not independently establish the construct validity of either the human or AI-generated marks~\cite{messick1995validity}. A marking process can be highly consistent yet systematically disagree with the reference, while a noisy human reference places a ceiling on the strength of any validity claim that can be made from AI-human agreement.

For marks used in ranking, progression, or grading, the property that matters is \emph{rank-order agreement} - whether a system places responses in the same quality order as the reference. This differs from \emph{distributional agreement}, which concerns only whether the overall spread of marks matches. A system can satisfy the second while failing the first entirely. If five essays have reference marks $\{12, 14, 16, 18, 20\}$, a marker who returns those same values but in reverse order reproduces the reference distribution exactly (identical mean and variance) yet achieves a Spearman correlation of $-1$, distributionally indistinguishable from the reference, but ordering the work backwards. Mean absolute error and calibration can be satisfied by distribution matching alone; only rank-order metrics test whether a system discriminates quality. Agreement with a human reference is therefore incomplete evidence of valid marking unless rank-order agreement is also established, and a reference that is itself unreliable cannot support strong claims about ranking at all.

\section{\label{sec:methods}Methods}
\subsection{\label{sec:overview}Overview}
This study evaluates contemporary large language models (LLMs) as automated marking tools across multiple physics assessment formats: structured exam questions, GCSE and A-Level problems with official reference solutions, university textbook questions, short-form physics essays, and scientific coding plots. The aim is not to benchmark answer-generation performance per se, but to assess the reliability, calibration, and validity of LLM-as-a-judge marking relative to human reference marks.

The datasets used here build directly on prior work exploring AI performance on physics examinations~\cite{curriculum-paper}, blinded human-AI comparisons in essay assessment~\cite{essays-paper} and blinded human-AI comparisons in scientific plots~\cite{coding-paper}. Rather than focusing on answer generation quality, the present study treats LLMs explicitly as evaluators. The use of LLM committees as aggregated evaluators follows recent evidence that a diverse panel of models reduces intra-model bias and outperforms single-model judges across multiple settings~\cite{verga2024poll}. Across all datasets, previously human-marked responses were re-evaluated via API calls to multiple LLMs under controlled prompting conditions. Within each task family, the same underlying response pool was marked across all models and experimental conditions, and prompts were held constant across providers apart from provider-specific formatting required by the API.

\subsection{\label{sec:metrics}Evaluation metrics}
Our two research questions both distinguish \emph{absolute accuracy} (how close the marks are to the human reference marks) from \emph{rank-order agreement} (whether responses are placed in the same quality order as human markers). These are separate properties as a system can match the distribution of human marks while failing to order responses by quality~\cite{messick1995validity,shankar2024validate_the_validators}. We therefore report metrics of both kinds throughout. Absolute accuracy is reported using Mean Absolute Error (MAE) and fractional MAE (MAE divided by available marks). Dividing by available marks places items with different mark totals on a common scale. These metrics measure the \emph{errors} part of RQ1 and the \emph{absolute accuracy} part of RQ2. Rank-order agreement is assessed independently of absolute accuracy using Spearman rank correlation ($\rho$), which measures whether the model orders responses by quality as humans do. We also report Quadratic Weighted Kappa (QWK)~\cite{cohen1968weighted_kappa}, which measures agreement on an ordered mark scale while correcting for chance agreement and penalising large mark disagreements more heavily than near misses. We report both because $\rho$ rewards correct ordering even under systematic harshness or leniency, whereas QWK is more sensitive to whether model and human marks land close together on the scoring scale. Agreement between these metrics guards against metric-dependent conclusions. These metrics address the rank-order and ordinal-agreement components of RQ1 and RQ2.

Where human marking reliability is reported, we use pairwise human-human MAE to quantify absolute disagreement between markers, and intraclass correlation coefficients (ICC2) to quantify the consistency of human marks across markers. These human-baseline metrics are important because agreement with a noisy human reference cannot be interpreted in the same way as agreement with a stable benchmark. For rank correlations on the essay dataset ($n = 55$ scripts), the effective sample size is small enough that $\rho$ values carry substantial uncertainty. Confidence intervals for Spearman $\rho$ are estimated by Fisher $z$-transformation: $z = \tfrac{1}{2}\ln\!\left(\tfrac{1+\rho}{1-\rho}\right)$, with a 95\% CI constructed as $z \pm 1.96/\sqrt{n-3}$ and then back-transformed. Because essay marks are awarded in 5-point increments, ties are common, which makes fine-grained rank separation harder and should be considered when interpreting small differences in $\rho$.

We use calibration curves~\cite{guo2017calibration_modern_nns} to assess score-scale alignment across the mark range. Their purpose is to distinguish systematic harshness or leniency across the score scale from failures of rank ordering: MAE averages over these effects, while Spearman $\rho$ is insensitive to a uniform score shift. These curves compare mean predicted fractional marks to mean human fractional marks within equal-width bins. Points close to the diagonal indicate that predicted and human marks are aligned on average within that region of the mark scale; systematic harshness or leniency appears as deviations below or above the diagonal. The committee mark is computed as the mean of the individual model marks. We treat the committee as an additional marking system, included to test whether aggregation across models recovers validity that individual models lack~\cite{verga2024poll}. Reporting both absolute-accuracy and rank-order metrics is essential because an LLM-as-a-judge system could achieve low MAE while still failing to distinguish good from poor work.

\subsection{\label{sec:question-datasets}Question Datasets}
For the structured physics questions, two datasets were used. Firstly, a set of university exam questions ($n = 771$) drawn from Durham University examinations (2018--2022), spanning advanced condensed matter physics, foundations modules, mathematical methods, astrophysics, cosmology, and atomic/optical physics. The solutions to these questions are not made public, hence they serve as a ``blinded'' dataset. Secondly, a set of curriculum questions was sourced from UK GCSE ($n=350$), A-Level ($n=370$), and university textbook questions ($n=431$), giving $n=1151$ structured questions in total. These questions and reference solutions are publicly available, hence they can be used to probe how access to a worked solution affects LLM marking. A full breakdown is shown in Table~\ref{tab:question-sources}. 

\begin{table*}[t]
\caption{\label{tab:question-sources}Question sources by dataset level. 
The curriculum subset (GCSE, A-Level, and textbook questions) contains 
$n = 1151$ items and the exam subset contains $n = 771$ items.}
\begin{ruledtabular}
\begin{tabular}{llld}
Level & Source & Years & \multicolumn{1}{c}{$n$} \\
\hline
Exam & Advanced Condensed Matter Physics & 2018--2022 & 154 \\
Exam & Foundations of Physics 1 (paper 1) & 2020--2022 & 26 \\
Exam & Foundations of Physics 2A & 2018--2022 & 112 \\
Exam & Foundations of Physics 3A & 2018--2022 & 110 \\
Exam & Mathematical Methods in Physics & 2018--2022 & 107 \\
Exam & Modern Atomic and Optical Physics 3 & 2020--2022 & 34 \\
Exam & Planets and Cosmology 3 & 2019--2022 & 61 \\
Exam & Theoretical Astrophysics & 2021--2022 & 19 \\
Exam & Theoretical Physics 2 & 2018--2022 & 109 \\
Exam & Theoretical Physics 3 & 2020--2022 & 39 \\
\hline
Textbook & Physics for Scientists and Engineers & -- & 179 \\
Textbook & Physics: Principles with Applications (7th ed.) & -- & 141 \\
Textbook & College Physics 2e & -- & 111 \\
\hline
A Level & OCR A-Level Physics A (papers 1, 2) & 2017--2020 & 243 \\
A Level & OCR A-Level Physics B (Advancing Physics) & 2020--2021 & 127 \\
\hline
GCSE & OCR GCSE Physics A (Gateway, papers 1--4) & 2018--2020 & 260 \\
GCSE & OCR GCSE Physics B (21st Century, papers 1, 3) & 2017--2022 & 90 \\
\hline
\textbf{Total} & & 2017--2022 & \textbf{1922} \\
\end{tabular}
\end{ruledtabular}
\end{table*}

The original question pools were manually inspected. Poorly formatted, ambiguous, or incomplete items (primarily from extracted exam \texttt{.tex} files) were removed to improve dataset integrity and ensure comparability across marking conditions. The exam questions were marked by their respective question setters ($n = 22$), which serve as the reference marks. The curriculum questions were marked by $n = 3$ human markers from Durham University taking a majority vote as the reference. These human markers were given the question and reference answer only, rather than the official marking schemes, to match the information that the LLMs were subsequently given.

Markers agreed unanimously on 76.7\% of items (883/1151) and a two-thirds majority was available for a further 20.5\% (236/1151). On the remaining 2.8\% (32/1151) all three markers assigned different marks; for these items, with no majority available, the reference was taken as the rounded mean of the three marks. Because curriculum items carry low mark totals (predominantly 1--2 marks), majority disagreements span a narrow range, and resolving the 32 no-majority items by rounded mean rather than by any single marker leaves all committee-level statistics unchanged to the precision reported. All responses were reused from a previous study comparing GPT-3.5 and GPT-4~\cite{curriculum-paper}, providing an existing human-marked response pool; the identity of the answering model is not a variable of interest here, as the present study concerns the behaviour of the LLM markers. The curriculum responses ($n = 1151$) were all generated by GPT-3.5, and the exam responses ($n = 771$) comprise 355 from GPT-3.5 and 416 from GPT-4.

\begin{figure}[t]
\centering
\includegraphics[width=\columnwidth]{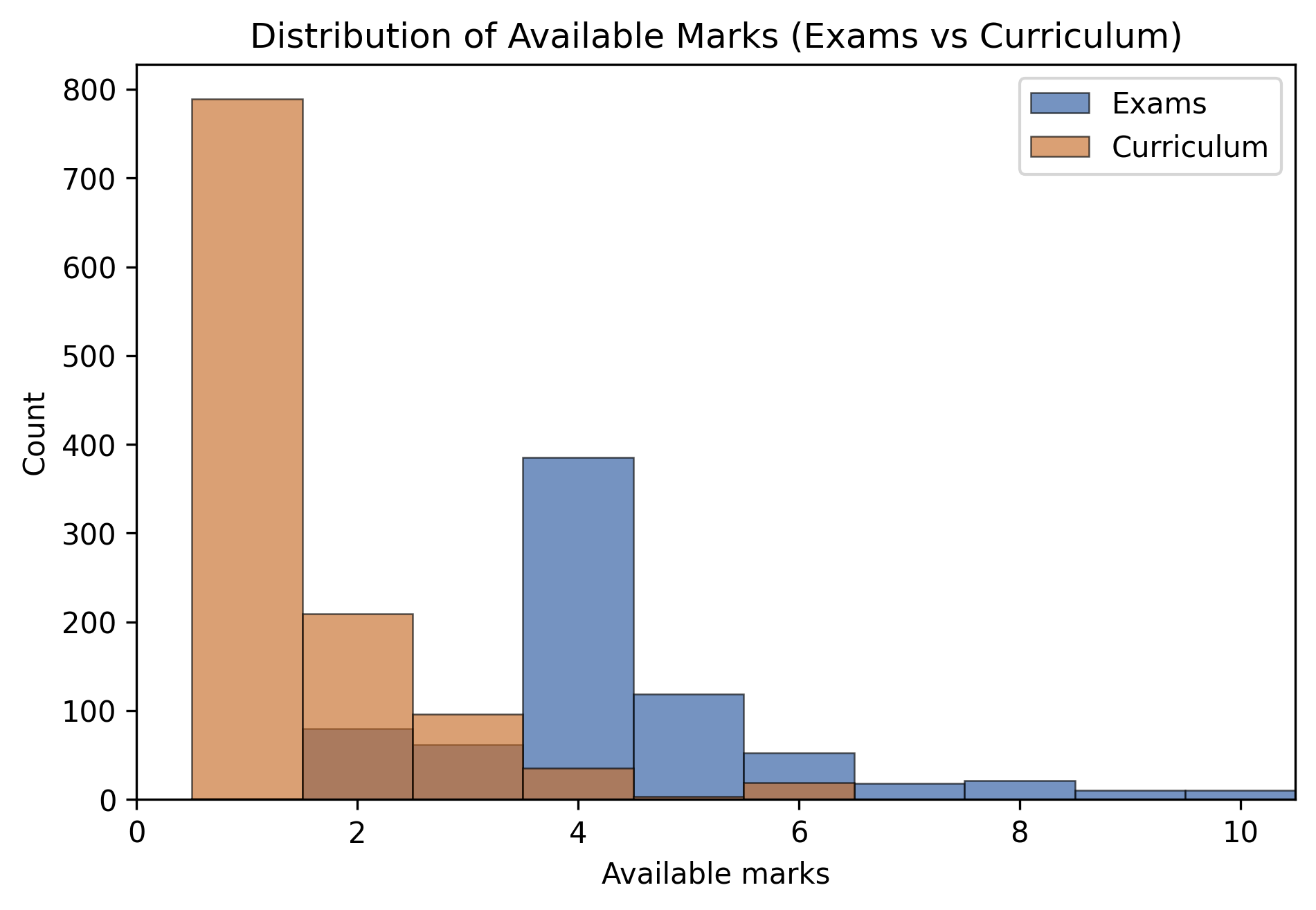}
\caption{\label{fig:exams-available}Distribution of marks available per question for the exams and curriculum datasets. Values above 10 marks (13/771 exam questions; 0 curriculum questions) are omitted for clarity.}
\end{figure}

These AI-generated responses achieved moderate overall scores against human marking providing a pool of both correct and incorrect responses. The mean per-question fractional score was $52.0\%$ for the exam dataset (marked by question setters) and $61.6\%$ for the curriculum dataset (majority vote of three markers), with $56.6\%$ of curriculum questions receiving full marks under the majority vote reference. A breakdown of the marks available per question is shown in FIG.~\ref{fig:exams-available}: curriculum questions are heavily concentrated at 1--2 marks per item (reflecting the short-answer format of GCSE and A-Level papers), whereas exam questions are more broadly distributed between 3 and 6 marks, reflecting the multi-step derivation style of university assessments. The higher mark allocations associated with university questions reflect a more complex rubric structure: multi-step derivations require the marker to assess representation, strategy, application, and logical progression jointly~\cite{docktor2016problem_solving_rubric}, a structure that a model marking without a scheme must infer from context alone.

A key pedagogical question for educators is how access to a reference solution influences LLM evaluations. Within individual institutions marking culture can vary widely~\cite{marshman2017contrasting_grading_approaches}; hence we used three marking regimes for the structured questions:
\begin{enumerate}
\item \textbf{Blinded marking.} The model was instructed to assess scientific correctness without access to a reference solution.
\item \textbf{With solution.} The model was instructed to compare the student response explicitly to the supplied solution and award marks strictly.
\item \textbf{With false solution.} To test for anchoring bias, models evaluated responses against a deliberately corrupted reference solution. This isolates whether the model independently verifies the physics or merely matches text patterns against the provided standard. False solutions were generated by requesting DeepSeek-V3.2 to apply one of five mechanical perturbations to the original answer: a factor-of-10 error, a sign flip, substitution of zero with a non-zero value, a multiple-choice swap, or doubling the main numerical value. The perturbation was required to produce an output visibly and unambiguously different from the correct answer while preserving surface formatting.
\end{enumerate}

A methodological caveat is warranted regarding the use of AI-generated responses as the student answer pool for the structured-question datasets. Because the Durham exam solutions are not distributed and cannot be reconstructed from student scripts under institutional data constraints, authentic student responses are unavailable for this setting. Further, the curriculum questions are from online sources and are not directly taught at Durham University meaning there isn't a pool of student answers to compare to. AI-generated answers were therefore used as a practical necessity rather than by design. These responses may exhibit systematically different error profiles compared with genuine student work. For example, AI answers may not approach problems using the methods taught on the course(s) which could plausibly inflate marking agreement by presenting models with responses closer to their own distributional expectations. The validity conclusions drawn from the structured-question datasets should therefore be interpreted as an upper bound on likely real-world performance. Where authorship effects can be tested directly (essays, scientific plots), we examine them in Section~\ref{sec:self-preference}.

\subsection{\label{sec:essay-dataset}Essay Dataset}
The essay dataset covered topics such as the Kuhn–Popper debate, interpretations of Quantum Mechanics and the historical development of scientific understanding of electricity. We initially used $n = 60$ scripts each containing five short-form essays (${\sim}300$ words each) graded holistically out of 100 in 5-point increments under standard UK classification bands. Because the five essays within each script were authored, submitted, and marked as a single submission, we chose a sample size of $n = 60$ rather than $n = 300$. To create a series of exemplars, we took five submissions from the 5th, 25th, 50th, 75th, and 95th percentile of aggregated human marks leaving a final dataset size of $n = 55$ scripts and $n = 275$ individual essays.

Element-level scores are reported for distributional description but all inferential statistics (rank correlations, confidence intervals) are computed at the script level. Evaluations were conducted by five independent markers with prior experience in the module who were blinded to the authorship of the work. The dataset includes both human-authored and AI-authored essays to increase the range of quality represented; authorship is treated as a data-collection covariate and is not a variable of interest in the present study, which concerns the behaviour of the AI markers rather than the AI authors.

To establish a baseline for human marking variability, we calculated pairwise inter-rater reliability across the five independent markers. The average human--human MAE was $6.84$ marks. Rank-order agreement between human markers was very limited, with an average pairwise Spearman correlation of just $\rho = 0.054$ and a single-rater Intraclass Correlation Coefficient (ICC2) of $0.035$ (95\% CI: $[-0.01, 0.11]$). Such low reliability is not unusual in short-form holistic marking, but it implies that fine-grained rank-ordering should be interpreted cautiously. This demonstrates that open-ended physics essay marking is an inherently noisy, highly subjective task where even expert human markers fail to reach consensus on rank-ordering - a baseline essential for interpreting all AI essay marking results that follow. This noisy baseline is not a methodological flaw, but an authentic representation of the reality of holistic grading in higher education. Apparent agreement with an aggregated human score can in principle be achieved by matching the distribution of a weakly reliable benchmark rather than by validly discriminating essay quality. For essay-level comparisons against a single human reference, we use the unrounded mean of the five human markers for each script as the aggregated human score. Three marking conditions were employed for the essays:

\begin{enumerate}
\item \textbf{Blind.} The model was instructed to assess scientific correctness without access to a mark scheme.
\item \textbf{With scheme.} The model was instructed to assess scientific correctness with use of the guidance scheme, which contains general content taught in the course and potential ideas to include. This is the method that the human markers used.
\item \textbf{With scheme and anchor.} The model was instructed to use the guidance scheme and a set of five exemplars drawn from the dataset at the 5th, 25th, 50th, 75th, and 95th percentile of average marks awarded by the human markers, each appended to the system prompt along with their scores. These 25 exemplar essays were removed from the evaluation set, leaving $n=275$ anchored-condition scores.
\end{enumerate}

\subsection{Qualitative disagreement analysis}
To address RQ2 beyond aggregate error metrics, we also examine cases in which the LLM markers unanimously disagree with the human reference. These cases are selected after quantitative marking using the criteria defined in Section~\ref{sec:qual-errors}, and are inspected to identify recurring mechanisms behind the disagreement, such as deference to a supplied reference or alternative but scientifically valid reasoning. The qualitative analysis is therefore used to interpret the mechanisms underlying the quantitative error patterns rather than as an independent sample of responses.

\subsection{\label{sec:plot-dataset}Scientific Plot Dataset}
We use $n = 1400$ graded scientific plots derived from 100 student and AI-generated submissions for the `Laboratory Skills and Electronics' module at Durham University, evaluated blindly by three independent markers. The dataset includes 14 distinct plot types per submission, covering physics topics such as numerical integration, Monte Carlo methods, and the Newton-Raphson method. The AI-generated plots were created with and without prompt engineering as covered in detail in~\cite{coding-paper}. Examples of plots produced by GPT-3.5 and GPT-4 are shown in Appendix~\ref{app-code-plots}. As the plots were submitted separately by the students we use $n = 1400$ as the sample size. Each plot was manually scored out of 5 based on its quality and accuracy in elucidating the underlying physics.

For this module there is no specific mark scheme for the plots. Human markers assessed the plots using a 0--5 scale of general plot quality: a score of 5 indicates a scientifically correct and clear plot with sensible axes, units, labels, and scale, while 0 indicates a blank, irrelevant, or unreadable plot. The LLMs were given the same 0--5 scoring guidance as the human markers. Importantly, this is not a test of unconstrained visual physics reasoning. Each plot occurred at the end of a Jupyter notebook task, and the model was shown the completed notebook content up to the point where the plotting code would begin, together with the final plot image. Human markers assessed the plots as presented in compiled PDF submissions, whereas the LLMs additionally received the preceding notebook context. This difference in available context should be considered when interpreting the human--LLM comparison.

\subsection{\label{sec:models}AI Models and Marking Conditions}
To capture the current state of the art, we use five models from leading labs: GPT-5.2 (OpenAI), Claude Opus 4.5 (Anthropic), Gemini 3 Pro (Google DeepMind), DeepSeek-V3.2 (DeepSeek), and Grok 4.1 (xAI). The core instruction prompt for structured questions is shown in FIG.~\ref{fig-prompt}. This instruction decomposes the evaluation into four explicit steps. Such structured prompting has been shown to improve zero-shot scoring accuracy in educational assessment settings~\cite{xiao2024cot_scoring}, and evaluation frameworks that incorporate intermediate reasoning (e.g.\ G-Eval) achieve stronger agreement with human graders than direct scoring approaches~\cite{liu2023geval}. In principle Chain of Thought prompting~\cite{cotpaper} or elsewise reasoning tokens can be used when generating responses, however given the different API structures the amount of reasoning used between models is difficult to keep constant. 

Pilot work on the curriculum dataset (Appendix~\ref{app-cot}) indicates that explicit Chain of Thought prompting can reduce fractional MAE for the same model (DeepSeek-V3.2) across blind and solution conditions, while the
DeepSeek-Reasoner model that generates native reasoning tokens prior to the final output does not yield comparable improvements and in some conditions performs worse than the baseline. Because different providers implement intermediate reasoning in architecturally distinct ways - explicit Chain of Thought in the prompt or reasoning tokens within the API calls - and because the volume of intermediate reasoning cannot be held constant across APIs, the main experiments use direct JSON-output prompting throughout to ensure a controlled cross-model comparison. The absolute
error values reported in the main text should therefore be interpreted as conservative estimates for structured questions under direct-output prompting.

\begin{figure}[t]
\centering
\begin{tcolorbox}[colframe=black!25, colback=blue!10]
\begin{verbatim}
INSTRUCTIONS:
1. Mark the student response by comparing 
it to the model solution.
2. Award a whole integer mark between 0 
and [AVAILABLE MARKS] (inclusive).
3. Do not invent marking criteria beyond 
what is implied by the solution.
4. If the student gives an alternative 
correct method not shown in the solution, 
award credit if it is clearly correct.

Return JSON exactly:
{"awarded_marks": integer}
\end{verbatim}
\end{tcolorbox}
\caption{\label{fig-prompt}Core marking prompt used for structured questions (solution condition).}
\end{figure}

For the solution and false solution conditions, the system message was replaced with an audit-marker instruction directing the model to mark by comparing the student response to the provided model solution and to be strict about what earns credit. The question block was extended to include a \texttt{MODEL SOLUTION} field alongside the student response and available marks. Temperature was set to 0 to reduce stochastic variation across repeated calls; pilot work, detailed in Appendix~\ref{app-temp}, suggested temperature changes did not affect the marks produced by the models. For essays, models were instructed to act as expert university markers using the same grading scale (0--100 in 5-point increments). For plots, models were instructed to act as expert markers using the same 0--5 scale; at the time of writing Grok 4.1 and DeepSeek-V3.2 did not accept image input, so only the remaining three models were used. Committee scores were computed as the mean of all individual model scores for each question - a transparent consensus estimator that preserves partial disagreement between models without introducing an additional fitted aggregation layer.

All marking was performed on the same response sets within each task family. Batch processing was used where supported by the provider API and elsewise requests were looped with standard exponential-backoff retries. All prompts used in this study are shown in Appendix~\ref{app-prompts}. In all settings, no post-processing was required to cap marks, as no model output exceeded the available maximum over all responses.

\section{\label{sec:results}Quantitative analysis of marking errors}
\subsection{\label{sec:structured}Structured Question Marking}
\subsubsection{Exams}
\begin{figure}[t]
\centering
\includegraphics[width=\columnwidth]{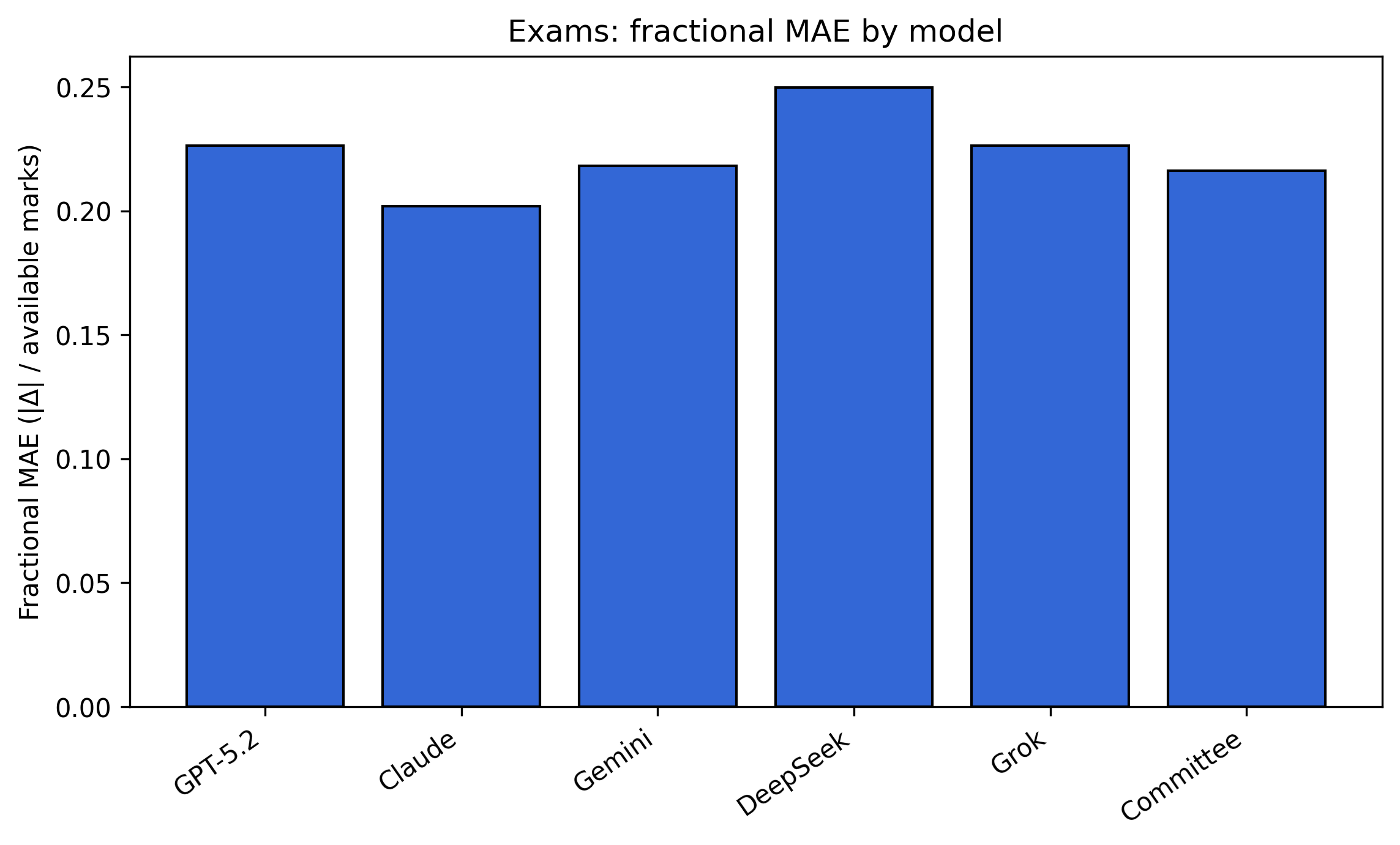}
\caption{\label{fig:exams-mae}Exams (Durham, blinded; $n=771$): fractional MAE by model. GPT-5.2: $0.226$; Claude: $0.202$; Gemini: $0.218$; DeepSeek: $0.250$; Grok: $0.226$; Committee: $0.216$.}
\end{figure}

Across the 771 Durham exam questions under blind marking, fractional MAE is tightly clustered across models (GPT-5.2: $0.226$, Claude: $0.202$, Gemini: $0.218$, DeepSeek: $0.250$, Grok: $0.226$, committee: $0.216$), with only modest separation between the best-performing model (Claude, $0.202$) and the weakest (DeepSeek, $0.250$) (FIG.~\ref{fig:exams-mae}). The committee aggregation ($0.216$) lies close to the model average, indicating that ensemble averaging provides limited additional benefit over the best individual model. These results are broadly consistent with prior feasibility work on LLM grading in introductory physics~\cite{kortemeyer2023toward}, while extending that work to a multi-model comparison across introductory-to-advanced undergraduate content under strictly blinded conditions. The rank-order agreement observed here is also consistent with prior findings that the structure and complexity of the assessment construct strongly influence the accuracy of automated scoring systems~\cite{zhai2022construct_scoring}.

Beyond absolute error, the models demonstrate robust rank-order agreement under blinded conditions (Table~\ref{tab:discriminative-validity-questions}). Rank correlations against human marks remain strong (e.g., Claude: Spearman $\rho = 0.690$, QWK $= 0.659$; GPT-5.2: $\rho = 0.646$, QWK $= 0.608$), indicating that models successfully rank student responses by physics quality even without reference solutions.

Across all conditions, absolute error shows a modest positive association with available marks (regression $R^2$: exams $0.199$, curriculum blind $0.205$, curriculum solution $0.210$), consistent with the idea that multi-step derivations require the marker to assess several co-dependent reasoning components simultaneously~\cite{docktor2016problem_solving_rubric}. Blinded automated marking may therefore be somewhat less reliable for longer-form, higher-mark questions, although available marks explain only a limited fraction of the total error variance.

\subsubsection{Curriculum questions}
\begin{figure*}[t]
\centering
\includegraphics[width=\textwidth]{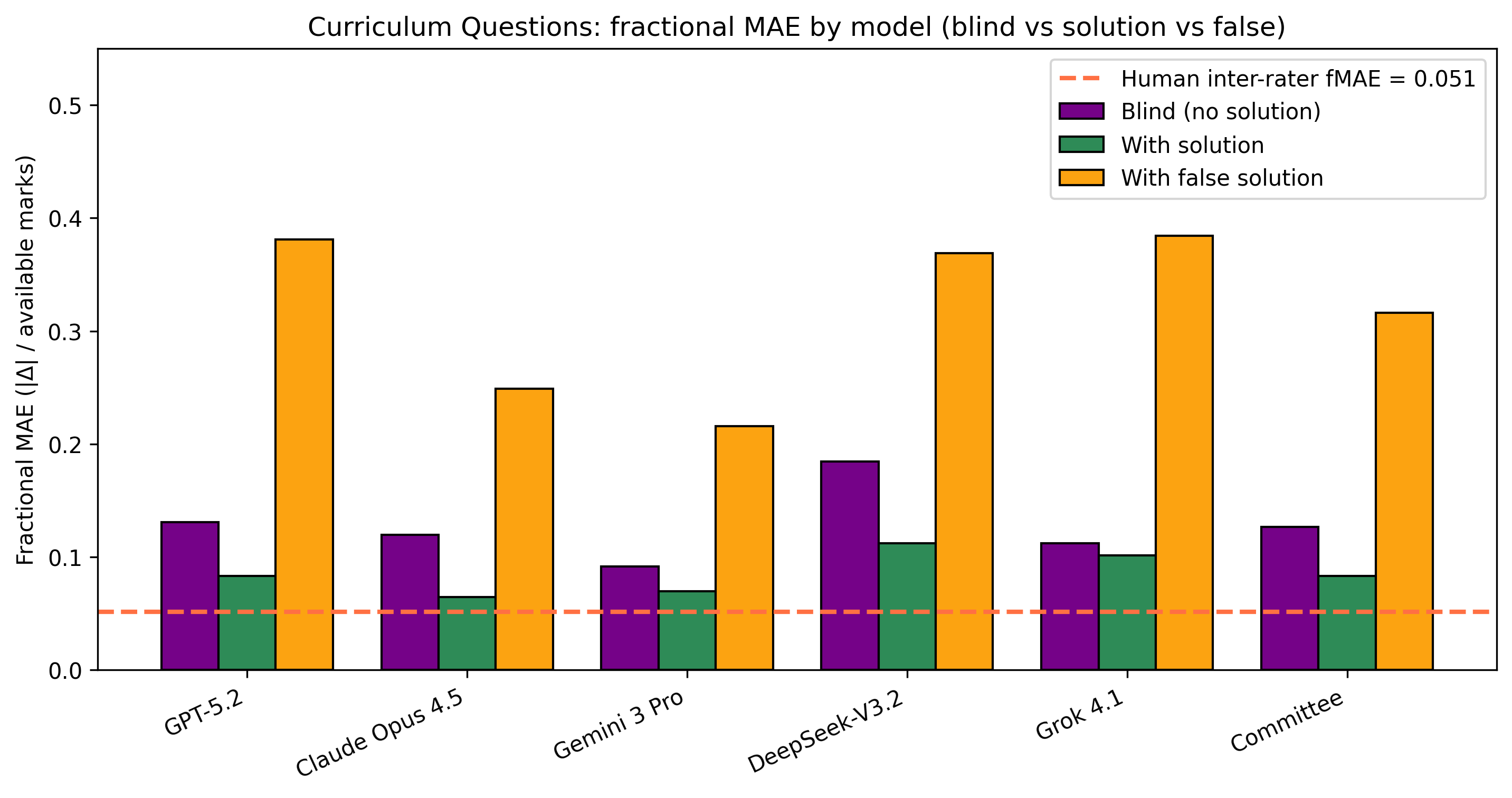}
\caption{\label{fig:curriculum-mae}Curriculum questions: fractional MAE by model (blind vs.\ solution vs.\ false solution) across GCSE ($n=350$), A-Level ($n=370$), and university textbooks ($n=431$). The human inter-rater fMAE ($n = 3$) is also shown.}
\end{figure*}

For GCSE, A-Level, and textbook questions (where reference solutions are available), providing the official solution consistently reduces fractional MAE across all models (GPT-5.2: $0.131\rightarrow0.083$; Claude: $0.120\rightarrow0.065$; Gemini: $0.092\rightarrow0.070$; DeepSeek: $0.185\rightarrow0.112$; Grok: $0.112\rightarrow0.101$; committee: $0.127\rightarrow0.083$) (FIG.~\ref{fig:curriculum-mae}). 

All of these values sit above the dashed reference line in FIG.~\ref{fig:curriculum-mae} at $0.051$, which shows the human inter-rater fractional MAE: the mean fractional absolute error of each human marker relative to the majority-vote reference across the curriculum questions. The non-zero value provides a human baseline for disagreement on this dataset.

Providing a false solution dramatically degrades absolute performance for all models, with fractional MAE rising to approximately $0.22$--$0.38$ (GPT-5.2: $0.381$; Claude: $0.249$; Gemini: $0.216$; DeepSeek: $0.369$; Grok: $0.384$; committee: $0.316$). This substantially exceeds the blind MAE for every model, confirming that the models attend to and defer to the provided reference rather than independently evaluating student responses on physical grounds. This deference to reference material mirrors experimental findings that LLMs exhibit statistically significant anchoring across providers~\cite{lou2024anchoring, nguyen2024anchoring, zheng2023mtbench_chatbot_arena_llm_judge}.

While the false solution significantly degrades absolute accuracy, it does not destroy rank-order agreement. Under blind and correct-solution conditions, models show strong alignment with human rankings (e.g., Gemini blind $\rho = 0.845$, QWK $= 0.860$; Claude solution $\rho = 0.885$, QWK $= 0.907$). When exposed to the false solution, these correlations dip but remain robust (e.g., Gemini false $\rho = 0.744$; Claude false $\rho = 0.719$; Table~\ref{tab:discriminative-validity-questions}). For structured mathematical physics, models retain the ability to rank good answers above bad answers while shifting their absolute scoring penalty to align with the incorrect reference.

\begin{table*}[t]
\caption{\label{tab:discriminative-validity-questions}Rank-order agreement (Spearman $\rho$ and QWK) for structured questions across conditions. 95\% CIs for $\rho$ by Fisher $z$-transformation.}
\begin{ruledtabular}
\begin{tabular}{lllccc}
Dataset & Model & Condition & $\rho$ & 95\% CI ($\rho$) & QWK \\
\hline
Exams & GPT-5.2 & Blind & 0.646 & $[0.603, 0.686]$ & 0.608 \\
 & Claude Opus 4.5 & Blind & 0.690 & $[0.651, 0.725]$ & 0.659 \\
 & Gemini 3 Pro & Blind & 0.634 & $[0.590, 0.675]$ & 0.627 \\
 & DeepSeek-V3.2 & Blind & 0.606 & $[0.559, 0.649]$ & 0.548 \\
 & Grok 4.1 & Blind & 0.640 & $[0.596, 0.680]$ & 0.637 \\
 & Committee & Blind & 0.696 & $[0.658, 0.731]$ & 0.650 \\
\hline
Curriculum & GPT-5.2 & Blind & 0.803 & $[0.781, 0.823]$ & 0.846 \\
 & Claude Opus 4.5 & Blind & 0.820 & $[0.800, 0.838]$ & 0.865 \\
 & Gemini 3 Pro & Blind & 0.845 & $[0.828, 0.861]$ & 0.860 \\
 & DeepSeek-V3.2 & Blind & 0.724 & $[0.696, 0.751]$ & 0.778 \\
 & Grok 4.1 & Blind & 0.821 & $[0.801, 0.839]$ & 0.852 \\
 & Committee & Blind & 0.849 & $[0.832, 0.865]$ & 0.867 \\
 & GPT-5.2 & Solution & 0.863 & $[0.847, 0.877]$ & 0.887 \\
 & Claude Opus 4.5 & Solution & 0.885 & $[0.872, 0.897]$ & 0.907 \\
 & Gemini 3 Pro & Solution & 0.873 & $[0.859, 0.886]$ & 0.877 \\
 & DeepSeek-V3.2 & Solution & 0.817 & $[0.797, 0.836]$ & 0.849 \\
 & Grok 4.1 & Solution & 0.828 & $[0.809, 0.846]$ & 0.858 \\
 & Committee & Solution & 0.885 & $[0.872, 0.897]$ & 0.898 \\
 & GPT-5.2 & False solution & 0.640 & $[0.605, 0.673]$ & 0.735 \\
 & Claude Opus 4.5 & False solution & 0.719 & $[0.690, 0.746]$ & 0.804 \\
 & Gemini 3 Pro & False solution & 0.744 & $[0.717, 0.769]$ & 0.817 \\
 & DeepSeek-V3.2 & False solution & 0.589 & $[0.550, 0.626]$ & 0.672 \\
 & Grok 4.1 & False solution & 0.619 & $[0.582, 0.653]$ & 0.721 \\
 & Committee & False solution & 0.772 & $[0.748, 0.794]$ & 0.758 \\
\end{tabular}
\end{ruledtabular}
\end{table*}

\begin{figure}[t]
\centering
\includegraphics[width=\columnwidth]{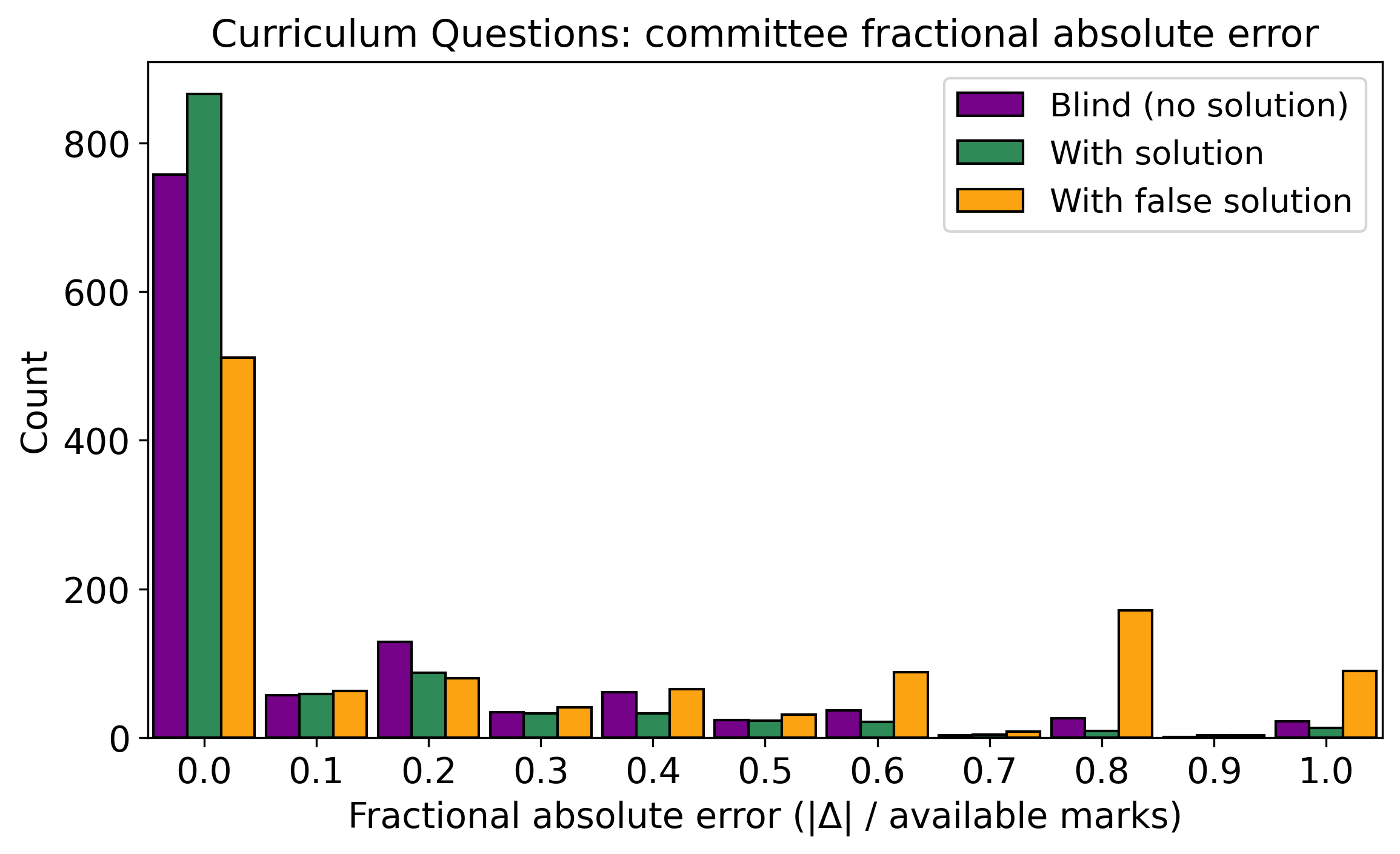}
\caption{\label{fig:curriculum-hist}Curriculum questions: distribution of committee fractional absolute error (blind vs.\ solution vs.\ false solution) across GCSE, A-Level, and textbook questions ($n=1151$ total).}
\end{figure}

The full distribution of committee absolute errors shifts left when the correct solution is provided, indicating systematic improvement rather than isolated correction of outliers (FIG.~\ref{fig:curriculum-hist}). Under the false solution condition the distribution shifts markedly right, with a pronounced secondary peak at fractional errors of $0.8$--$1.0$, corresponding to cases where the model awarded near-zero marks because the student's correct answer differed substantially from the deliberately incorrect reference.

\begin{figure}[t]
\centering
\includegraphics[width=\columnwidth]{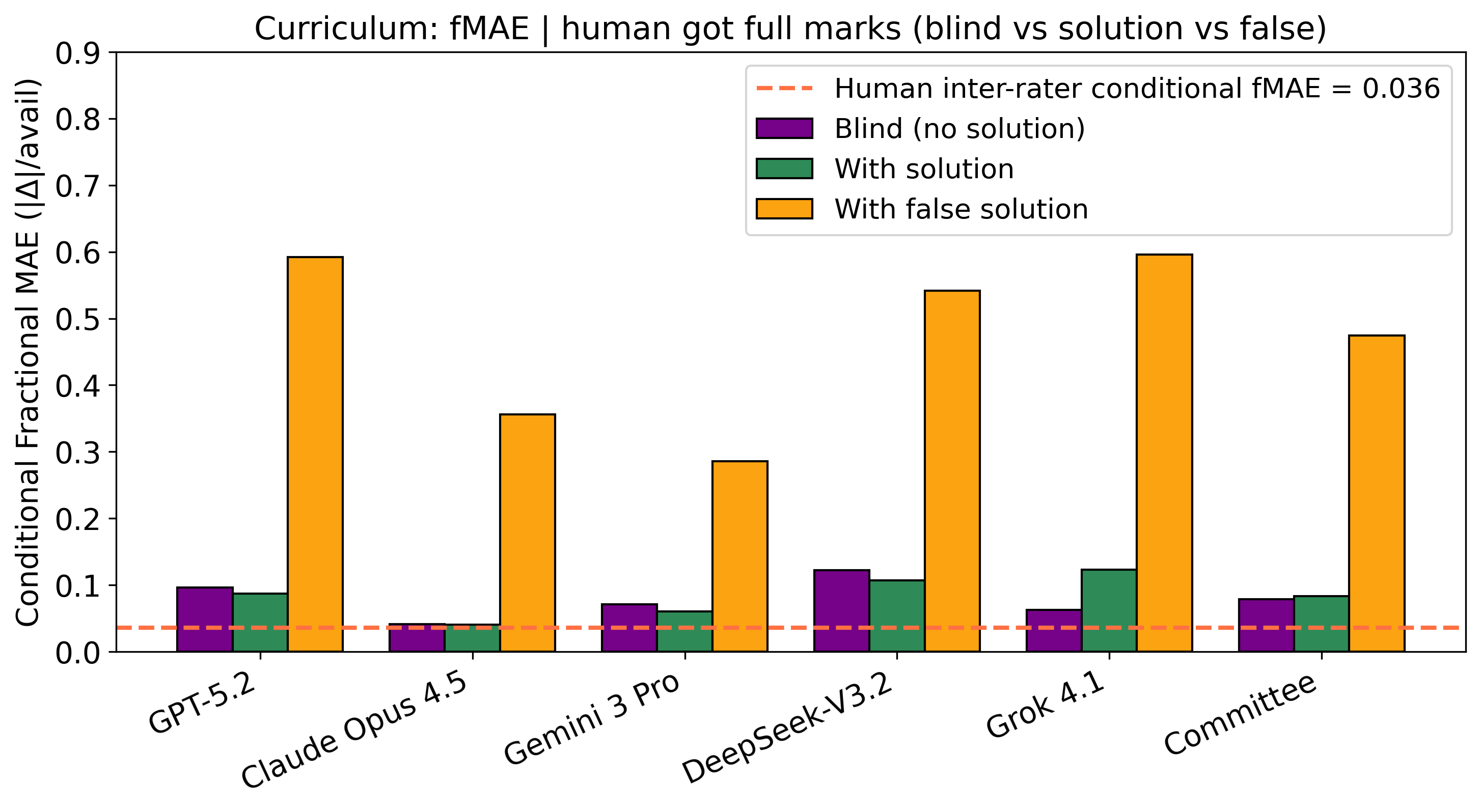}
\caption{\label{fig:fullmarks}Curriculum questions: conditional fractional MAE for questions on which the human awarded full marks ($n=652$). All models maintain low MAE under blind and solution conditions (${\sim}0.04$--$0.12$) but suffer large errors under the false solution (${\sim}0.29$--$0.60$).}
\end{figure}

To probe whether the effect of solution quality is concentrated on the hardest questions or is more uniform, FIG.~\ref{fig:fullmarks} shows the conditional fractional MAE restricted to the $n=652$ questions on which the human marker awarded full marks. In this subset, blind and solution-provided performance are broadly comparable (committee: $0.079$ blind, $0.084$ solution), indicating that for questions with unambiguously correct student responses, models are already close to ceiling under blind conditions. Under the false solution condition, however, conditional MAE rises sharply (committee: $0.474$), confirming that the effect is driven by active misdirection from the incorrect reference. Notably, the best-performing individual models under blind and solution conditions (Claude blind: $0.041$; Claude solution: $0.041$) outperform the human inter-rater conditional baseline on the full-marks subset, while the committee ($0.079$ blind; $0.084$ solution) sits modestly above it. The false solution condition exceeds the human baseline by a large margin for all models.

\begin{figure}[t]
\centering
\includegraphics[width=\columnwidth]{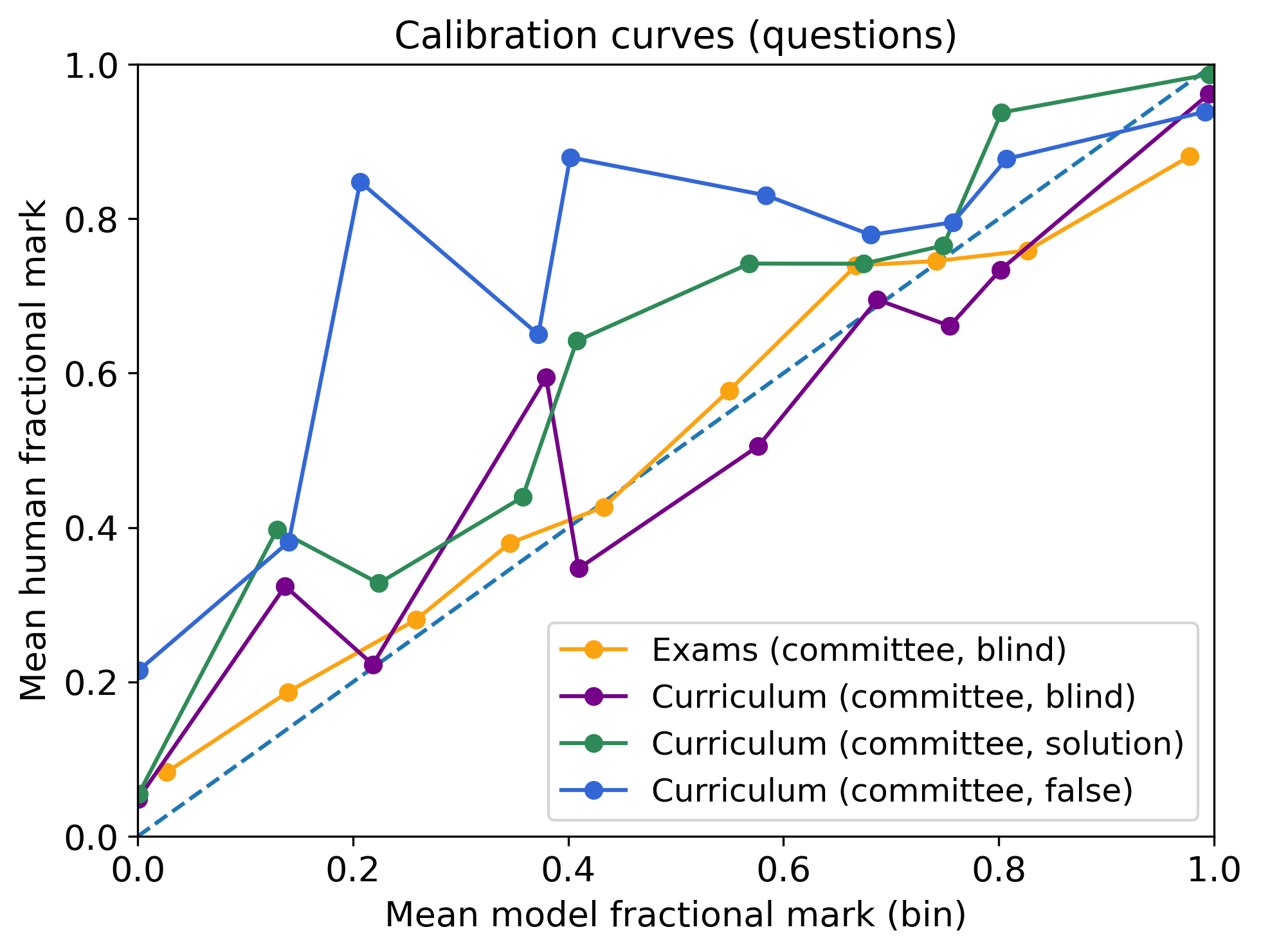}
\caption{\label{fig:questions-calibration}Calibration curves for structured questions (fractional marks) across exams and curriculum conditions. The dashed diagonal represents perfect score-scale alignment.}
\end{figure}

Calibration curves across exams and curriculum questions lie close to the diagonal for the blind and solution cases with modest deviations in mid-to-high mark bins under blind conditions that are reduced when solutions are provided (FIG.~\ref{fig:questions-calibration}). Providing a false solution causes large deviations from the diagonal.

\subsection{\label{sec:essays}Essay Marking}

\subsubsection{Mark distributions}
\begin{figure}[t]
\centering
\includegraphics[width=\columnwidth]{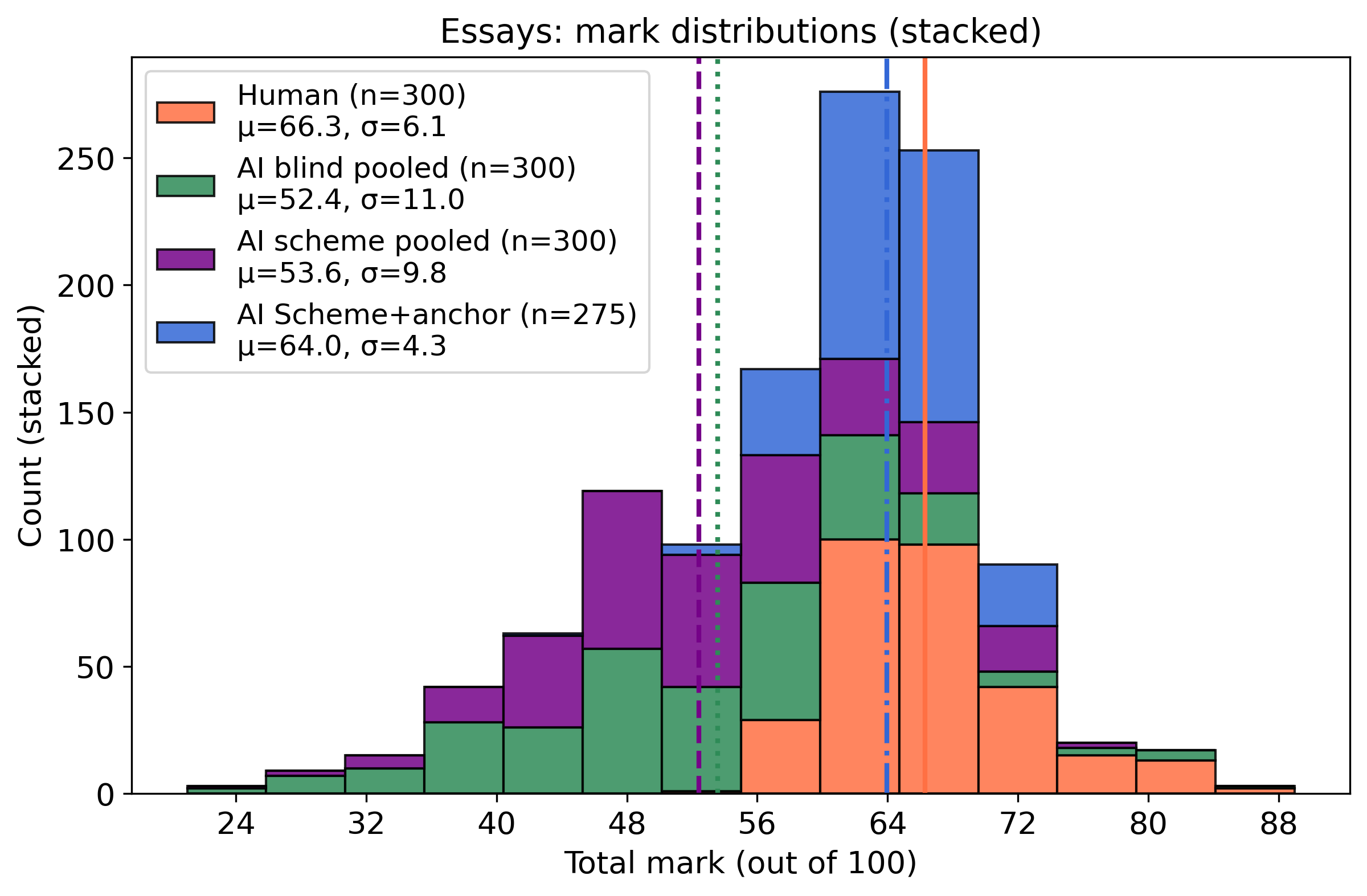}
\caption{\label{fig:essay-hist}Essay mark distributions (stacked; $n=300$ per condition except scheme+anchor where $n=275$): humans ($\mu=66.3$, $\sigma=6.1$), AI blind pooled ($\mu=52.4$, $\sigma=11.0$), AI scheme pooled ($\mu=53.6$, $\sigma=9.8$), AI scheme+anchor pooled ($\mu=64.0$, $\sigma=4.3$). Vertical lines indicate pooled means for each condition.}
\end{figure}

FIG.~\ref{fig:essay-hist} shows stacked distributions of human marks ($n = 300$ marker-level scores, mean $66.3$, $\sigma=6.1$), AI blind pooled (mean $52.4$, $\sigma=11.0$), AI scheme pooled (mean $53.6$, $\sigma=9.8$), and AI scheme+anchor pooled ($n = 275$, mean $64.0$, $\sigma=4.3$). The reduced $n$ in the anchored condition reflects the deliberate removal of the five exemplar scripts used for anchoring calibration, which were selected at the 5th, 25th, 50th, 75th, and 95th percentiles of the human mark distribution. All per-script inferential statistics (Spearman $\rho$, QWK, MAE, and calibration curves) are computed over the matched $n = 55$ scripts throughout, ensuring like-for-like comparisons across conditions; the $n = 300$ pooled distributions are shown for illustrative purposes only.

The human distribution is approximately unimodal, centred around $65$--$70$, with limited spread. Blind and scheme-only AI distributions are both shifted approximately 14 marks below the human mean and display considerably inflated variance. Adding the scheme alone provides negligible improvement: the scheme-only mean ($53.6$) is only 1.2 marks above the blind mean ($52.4$), and the variance reduction is modest (from $\sigma=11.0$ to $\sigma=9.8$). By contrast, the scheme+anchor condition qualitatively transforms the distribution: the mean rises to $64.0$ (within 2.3 marks of the human mean) and variance is compressed to $\sigma=4.3$, below even the human standard deviation of $6.1$.

This distributional shift should not be read straightforwardly as improved essay evaluation. Because the human benchmark itself has very low inter-rater reliability, close matching of its overall mean and variance can in principle be achieved by learning the expected score distribution without learning to discriminate essay quality. Distributional agreement can occur in a low-reliability task.

\subsubsection{Essay MAE by model}

\begin{figure*}[ht]
\centering
\includegraphics[width=\textwidth]{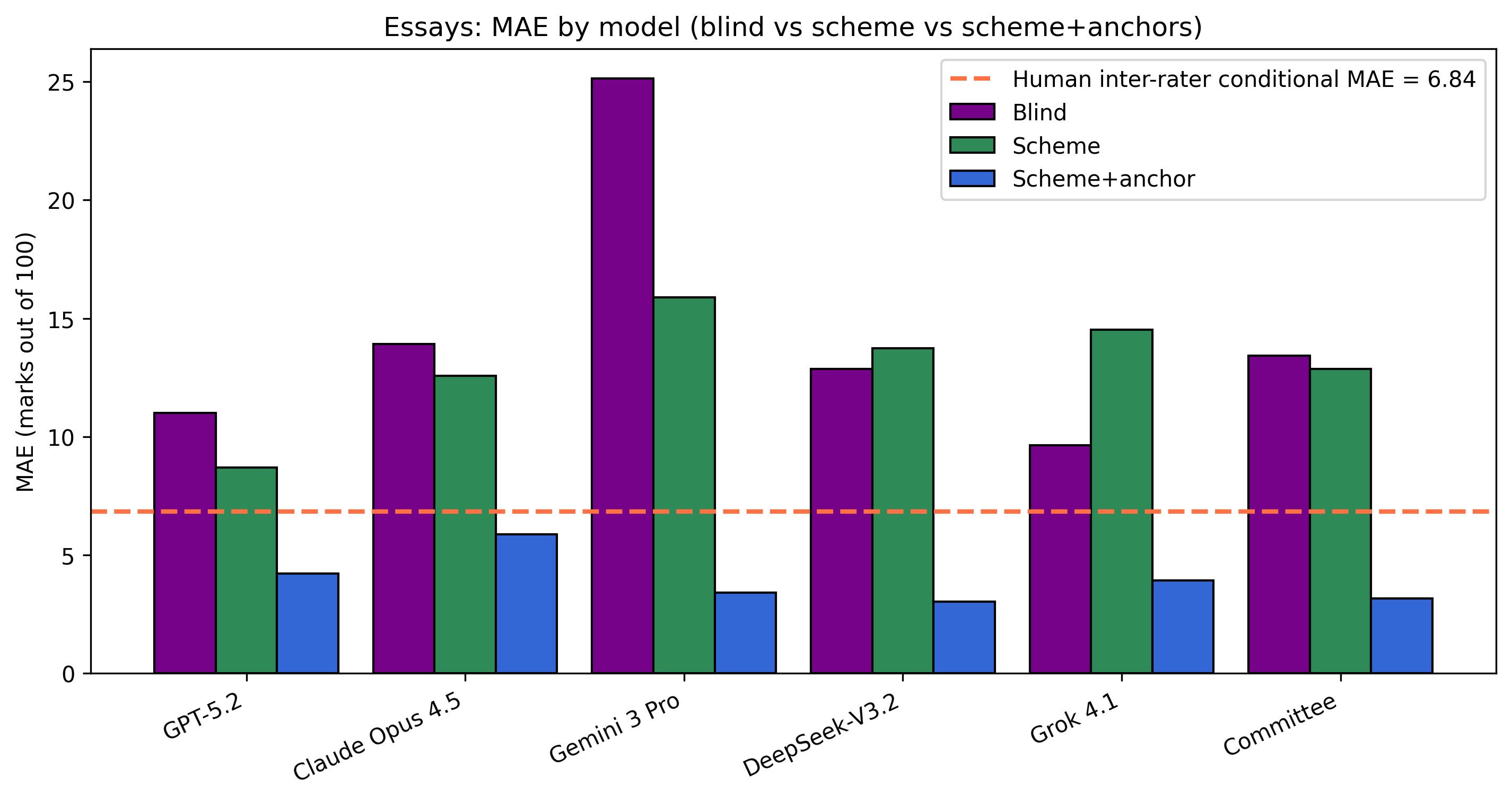}
\caption{\label{fig:essay-mae}Essays: MAE by model. Committee MAE falls from 13.42 (blind) and 12.86 (scheme) to 3.16 with anchoring. The dashed line indicates the average human--human MAE of 6.84.}
\end{figure*}

Model-level MAE (marks out of 100) across blind, scheme, and scheme+anchor conditions is shown in FIG.~\ref{fig:essay-mae}. Under blind conditions, MAE ranges from $9.6$ (Grok) to $25.1$ (Gemini) - a factor of more than $2.5$ across models - indicating substantial model-to-model variation in default scoring leniency. Adding the scheme reduces MAE for GPT-5.2 ($11.0\rightarrow8.7$) and Claude ($13.9\rightarrow12.6$) but actually increases it for Grok ($9.6\rightarrow14.5$) and DeepSeek ($12.9\rightarrow13.7$), suggesting that a content-based scheme without distributional anchoring can shift some models toward different systematic biases. The scheme+anchor condition, by contrast, produces consistent and dramatic improvements across all models: MAE falls to $4.2$ (GPT-5.2), $5.9$ (Claude), $3.4$ (Gemini), $3.0$ (DeepSeek), $3.9$ (Grok), and $3.2$ for the committee, representing reductions of roughly $3$--$8\times$ relative to blind marking. This wide inter-model variation is consistent with automated essay scoring findings showing that default LLM scoring behaviour reflects substantial leniency and variance shifts across model and task framing~\cite{pack2024llm_aes_validity_reliability,floden2025grading_exams_chatgpt}. A broader reflection on the automated essay scoring field argues that benchmark-focused evaluation has systematically neglected construct validity and generalisation~\cite{li2024aes_reflection}, while rubric-based LLM essay grading can achieve high reliability when rubrics are sufficiently granular~\cite{yavuz2025rubric_essay}.

These MAE values must be interpreted against the human baseline. The average human--human MAE of $6.84$ marks means that the blind and scheme-only AI conditions (committee MAE $\approx 13.4$) are roughly double the average disagreement between expert human markers. Conversely, the anchored condition's committee MAE of $3.16$ marks is tighter than human markers are capable of achieving against each other. Rather than being evidence of superior marking, this indicates that anchored prompts can reduce error against a noisy aggregated benchmark by pulling model scores toward the centre of the observed human distribution.

\begin{figure*}[t]
\centering
\includegraphics[width=\textwidth]{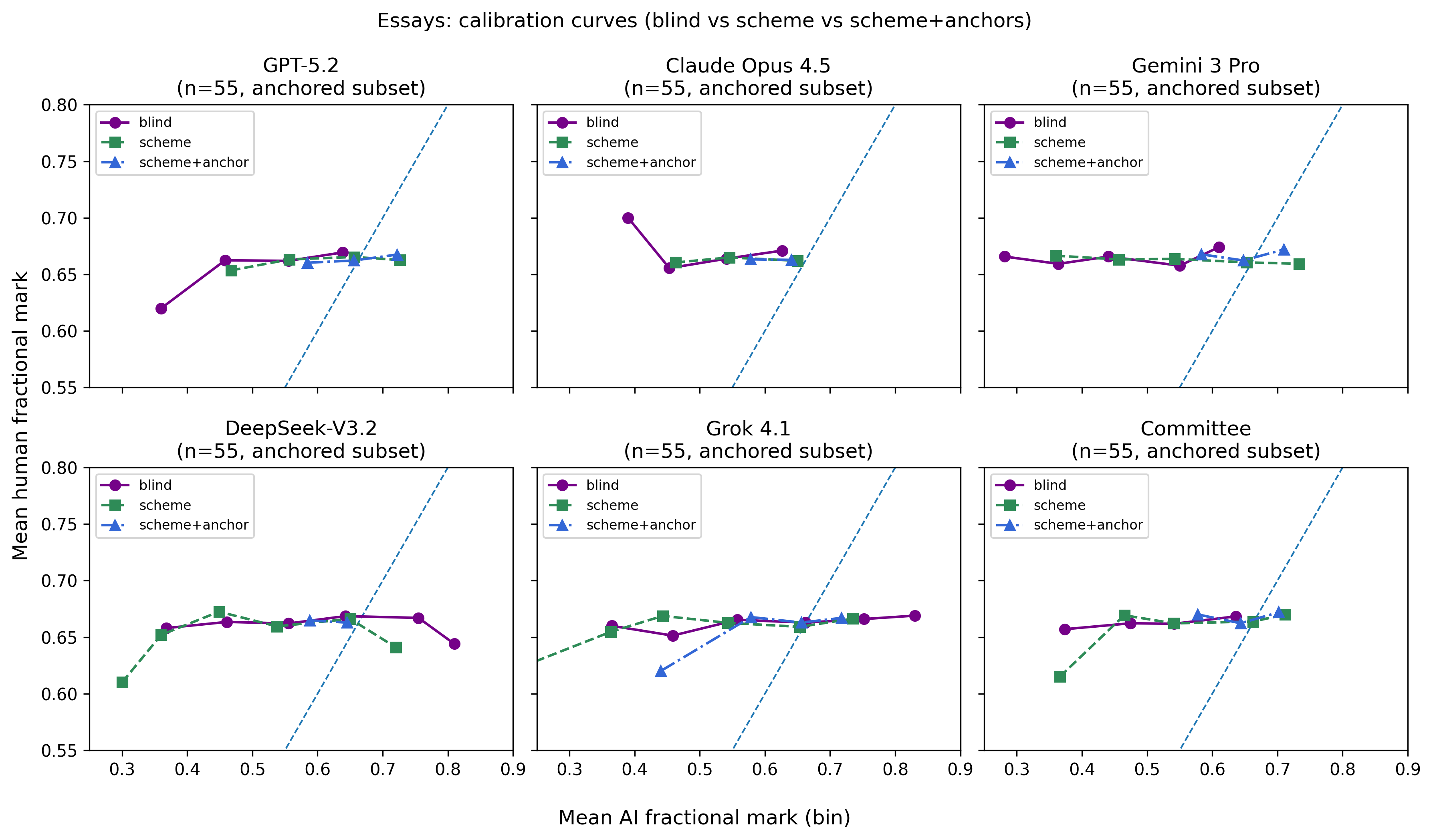}
\caption{\label{fig:essay-calibration}Essays: calibration curves by model and condition (blind vs.\ scheme vs.\ scheme+anchor; scheme+anchor uses $n=275$). The dashed diagonal represents perfect score-scale alignment. Blind and scheme-only conditions show systematic under-marking relative to human scores; anchoring improves score-scale alignment without recovering rank-order agreement.}
\end{figure*}

Calibration curves confirm that blind and scheme-only marking are systematically harsh relative to human markers (FIG.~\ref{fig:essay-calibration}): points fall substantially below the diagonal, indicating that when a model assigns a given fractional mark, the corresponding human fractional mark is consistently higher. As with the MAE results, however, improved score-scale alignment should not be equated with improved validity unless the system also preserves the ability to distinguish stronger from weaker essays.

\subsubsection{Rank-order agreement and anchoring bias}
While anchored exemplars dramatically improve distributional alignment and MAE, rank-order agreement remains absent throughout. We computed Spearman rank correlation ($\rho$) and QWK against the aggregated human scores (Table~\ref{tab:discriminative-validity-essays}).

\begin{table*}[t]
\caption{\label{tab:discriminative-validity-essays}Essay rank-order agreement (Spearman $\rho$ and QWK) by model and condition ($n=55$ scripts). 95\% CIs for $\rho$ by Fisher $z$-transformation. Human--human average $\rho = 0.054$ (95\% CI $[-0.21, 0.31]$) and ICC2 $= 0.035$ (95\% CI $[-0.01, 0.11]$).}
\begin{ruledtabular}
\begin{tabular}{llccc}
Model & Condition & $\rho$ & 95\% CI ($\rho$) & QWK \\
\hline
Human avg. &  -  & 0.054 & $[-0.21, 0.31]$ &  -  \\
\hline
GPT-5.2 & Blind & 0.109 & $[-0.15, 0.35]$ & 0.029 \\
GPT-5.2 & Scheme & 0.006 & $[-0.25, 0.26]$ & 0.020 \\
GPT-5.2 & Scheme+anchor & $-0.021$ & $[-0.27, 0.23]$ & 0.023 \\
Claude Opus 4.5 & Blind & 0.154 & $[-0.10, 0.39]$ & 0.020 \\
Claude Opus 4.5 & Scheme & 0.049 & $[-0.21, 0.30]$ & 0.008 \\
Claude Opus 4.5 & Scheme+anchor & $-0.016$ & $[-0.27, 0.24]$ & 0.012 \\
Gemini 3 Pro & Blind & $-0.013$ & $[-0.27, 0.25]$ & 0.003 \\
Gemini 3 Pro & Scheme & $-0.103$ & $[-0.35, 0.16]$ & $-0.011$ \\
Gemini 3 Pro & Scheme+anchor & 0.112 & $[-0.15, 0.36]$ & 0.091 \\
DeepSeek-V3.2 & Blind & 0.091 & $[-0.17, 0.34]$ & 0.023 \\
DeepSeek-V3.2 & Scheme & $-0.049$ & $[-0.30, 0.21]$ & 0.002 \\
DeepSeek-V3.2 & Scheme+anchor & 0.016 & $[-0.25, 0.28]$ & 0.045 \\
Grok 4.1 & Blind & 0.090 & $[-0.17, 0.34]$ & 0.048 \\
Grok 4.1 & Scheme & $-0.024$ & $[-0.28, 0.23]$ & 0.002 \\
Grok 4.1 & Scheme+anchor & 0.034 & $[-0.22, 0.28]$ & 0.052 \\
Committee & Blind & 0.088 & $[-0.17, 0.34]$ & 0.019 \\
Committee & Scheme & $-0.044$ & $[-0.29, 0.21]$ & 0.002 \\
Committee & Scheme+anchor & 0.034 & $[-0.22, 0.28]$ & 0.066 \\
\end{tabular}
\end{ruledtabular}
\end{table*}

Even without anchoring, LLM essay rank-order agreement is already poor. Under blind conditions, rank correlations are weak and near-zero (e.g., Claude Opus 4.5 blind: $\rho = 0.154$, 95\% CI $[-0.10, 0.39]$; GPT-5.2 blind: $\rho = 0.109$; Committee blind: $\rho = 0.088$). A correlation of $\rho \approx 0.1$ is insufficient for a marking system intended to distinguish student performance: the model is only marginally better than chance at ordering essays by quality relative to the aggregated human benchmark. Adding the scheme does not improve rank-order agreement: all scheme-condition $\rho$ values remain near-zero and statistically indistinguishable from the blind values given the width of the confidence intervals (e.g., GPT-5.2 scheme: $\rho = 0.006$, 95\% CI $[-0.25, 0.26]$; committee scheme: $\rho = -0.044$, 95\% CI $[-0.29, 0.21]$). The anchoring condition shows a similar pattern: committee $\rho$ sits at $0.034$ (95\% CI $[-0.22, 0.28]$), with individual models GPT-5.2 ($\rho = -0.021$) and Claude ($\rho = -0.016$) likewise consistent with zero. Across all three conditions, no model or committee achieves a $\rho$ value whose confidence interval excludes zero, and the between-condition differences are far smaller than the within-condition uncertainty.

These metrics reveal a more subtle failure mode than simply ``AI is bad at essays.'' The human benchmark itself exhibits extremely weak rank reliability, so this essay format is already a poor instrument for stable single-rater ordering. Anchored exemplars allow the models to move much closer to the aggregated human score distribution while rank-order agreement remains absent throughout. Anchoring reduces error against a noisy consensus-derived reference without solving the underlying validity problem; the models behave less like independent evaluators and more like systems inferring where marks are expected to lie on average.

\subsection{\label{sec:plots}Code Plot Marking}
\begin{figure}[ht]
\centering
\includegraphics[width=\columnwidth]{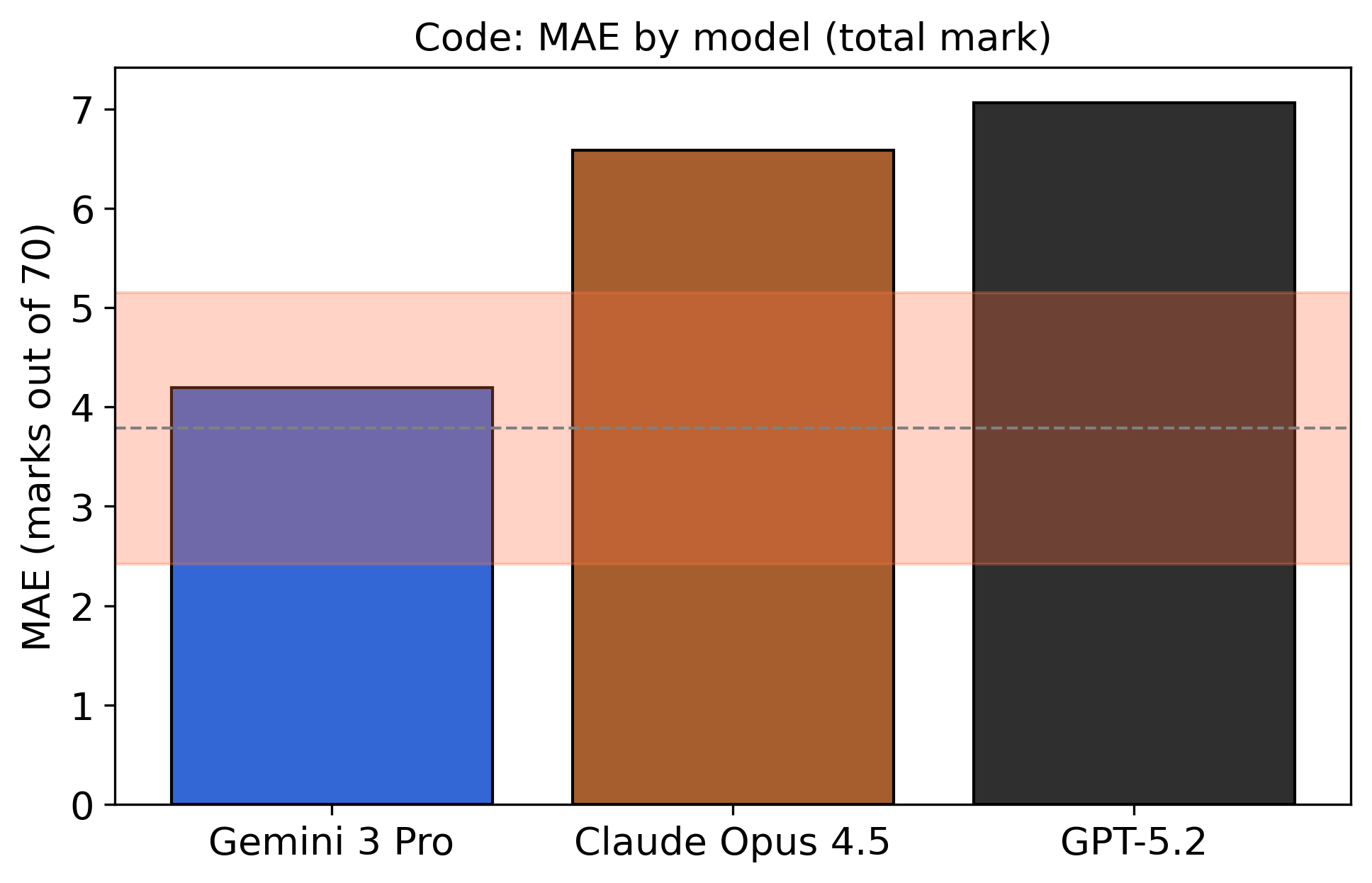}
\caption{\label{fig:code-mae}Scientific plots: MAE by model (total mark out of 70) across $n=1400$ plot elements (100 scripts). Human--human inter-rater MAE ranges from $2.42$ to $5.15$, providing a reference scale.}
\end{figure}

For structured code-based plotting assignments (total out of 70 marks, aggregated across 14 plots per script), MAE values differ meaningfully across models: Gemini is the strongest performer ($4.2$ marks), followed by Claude ($6.6$) and GPT-5.2 ($7.1$) (FIG.~\ref{fig:code-mae}). Human--human inter-rater MAE ranges from $2.42$ (markers 1 vs.\ 3) to $5.15$ (markers 2 vs.\ 3). Gemini's MAE therefore falls within the range of human--human disagreement - a threshold psychometric approaches to AI-assisted grading identify as meaningful for conditional deployment~\cite{kortemeyer2024psychometrics}.

A detailed characterisation of the qualitative features that distinguish AI-generated from student-authored plots in this dataset including systematic differences in colour scheme, axis alignment, and layout is provided in prior work~\cite{coding-paper}, where the same submissions were evaluated for authorship detectability. The present results extend that analysis by showing that these surface differences do not substantially impair AI marking validity: despite the stylistic differences between student and AI-generated code, rank-order agreement remains high ($\rho > 0.83$ across all models), suggesting the models are responding to features of plot quality described by the scoring scale rather than authorship-correlated stylistic cues.

\begin{figure}[ht]
\centering
\includegraphics[width=\columnwidth]{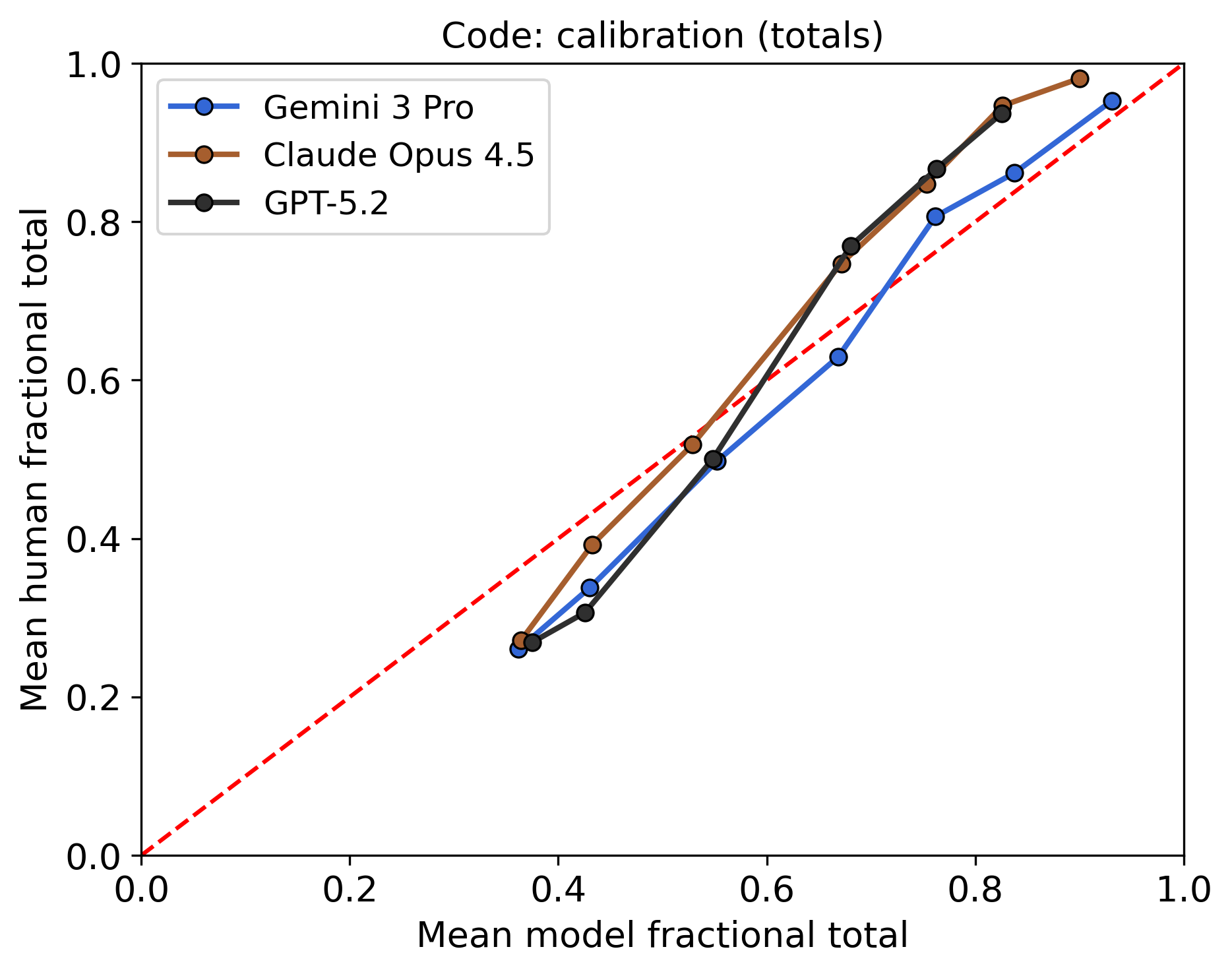}
\caption{\label{fig:code-cal}Scientific plots: calibration (total fractional mark) across $n=1400$ plot elements (100 scripts). The dashed diagonal represents perfect score-scale alignment.}
\end{figure}

Rank-order agreement on scientific plots is exceptionally strong - and contrasts sharply with the essay results. Rank correlations indicate tight alignment with human ranking preferences (Gemini: $\rho = 0.881$, QWK $= 0.926$; Claude: $\rho = 0.909$, QWK $= 0.802$; GPT-5.2: $\rho = 0.838$, QWK $= 0.725$). These values are several times higher than those observed in essay marking under any condition, confirming that the models genuinely differentiate between good and poor visual data representation rather than merely matching a central score tendency. Calibration curves show close adherence to the diagonal for all three models across the central fractional range (FIG.~\ref{fig:code-cal}), with some overestimation in the lowest-score bin and near-perfect agreement in the upper range.

This strong performance should still be interpreted narrowly. The task here is not arbitrary graph interpretation or deep multimodal scientific reasoning, but contextualised judgement of plot quality within a constrained Jupyter notebook-based workflow. The result supports LLM/VLM use for scale-guided evaluation of scientific plots in this specific setting, not a broader claim that visual physics assessment is generally solved. GPT-4V has separately been shown to score student-drawn scientific models with moderate accuracy using structured prompting~\cite{lee2025nerif}, and benchmarking on scientific plot comprehension confirms that current multimodal LLMs achieve meaningful but imperfect understanding of standard matplotlib-style outputs~\cite{wu2025plot2code}. The strong performance observed here is therefore best understood as reflecting the constrained nature of the task rather than general visual physics reasoning.

\begin{figure}[t]
\centering
\includegraphics[width=\columnwidth]{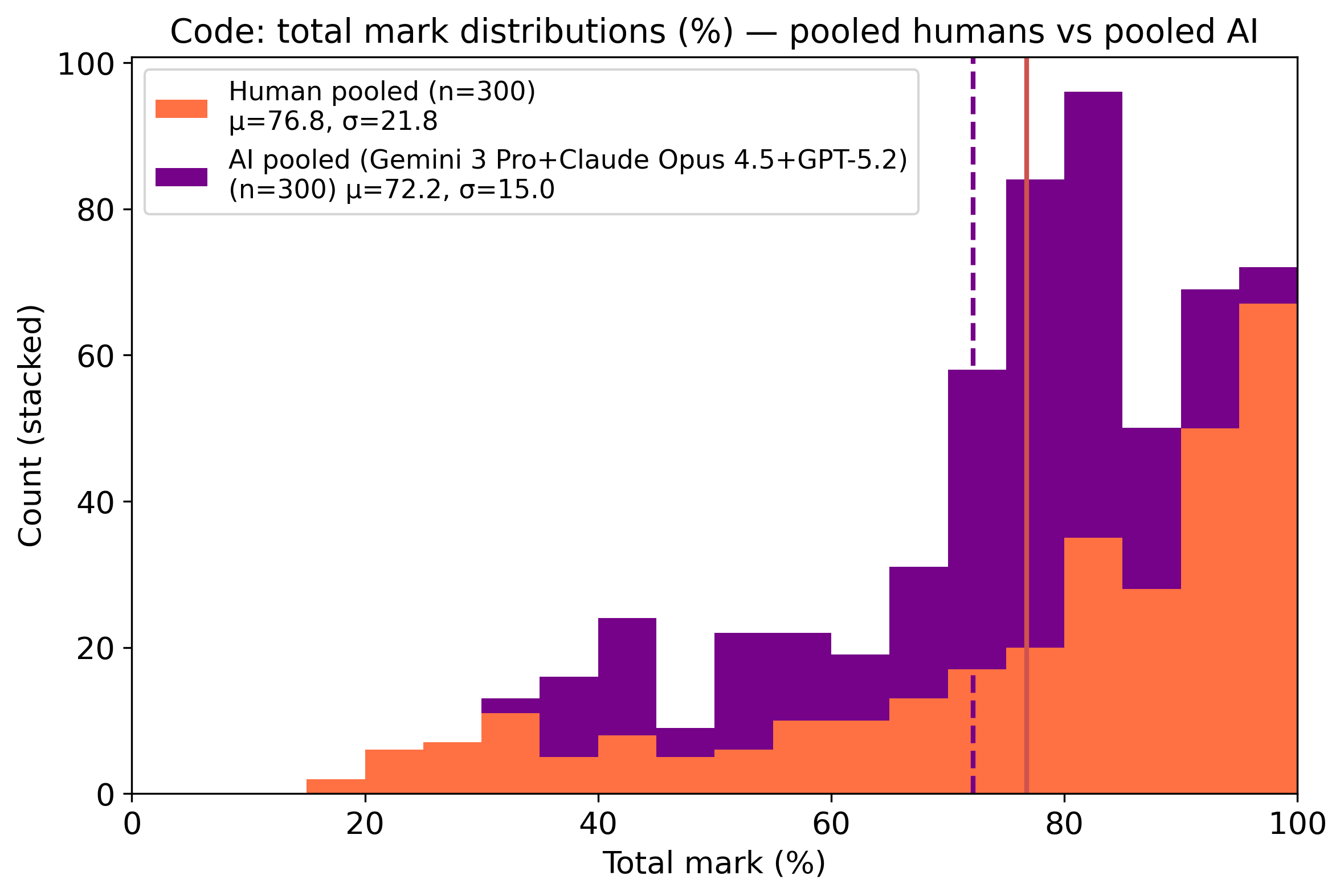}
\caption{\label{fig:code-hist}Scientific plots: total mark distributions (\%) - pooled humans ($\mu=76.8\%$, $\sigma=21.8$; $n=300$) vs.\ pooled AI ($\mu=72.2\%$, $\sigma=15.0$; $n=300$). AI marking is slightly more conservative and exhibits reduced variance relative to human markers.}
\end{figure}

The pooled histogram (FIG.~\ref{fig:code-hist}) shows that human pooled scores (mean $76.8\%$, $\sigma=21.8$) span a broader range than AI pooled scores (mean $72.2\%$, $\sigma=15.0$). The AI distribution is more concentrated around the $75$--$85\%$ band, consistent with a systematic tendency toward conservative central marking. The AI mean (72.2\%) is 4.6 percentage points lower than the human mean (76.8\%). Human markers differ by 2.42–5.15 marks on average ($\approx$3.5–7.4\% of the total score), indicating that the AI–human difference lies within the range of typical human grading disagreement.

\subsection{\label{sec:self-preference}Authorship effects}
The structured-question responses evaluated in this study are AI-generated rather than human-authored. For the exam questions, this is due to assessment integrity and for the curriculum questions this is a result of the question content not being directly taught at the institution the research took place at: Durham University. To assess whether this affects the reliability of our validity conclusions, we examined authorship effects in the datasets where student answers were available: the essay and code datasets. In the essay dataset - where human markers assigned near-identical mean scores to human- and AI-authored work ($\mu = 66.2$ vs.\ $66.4$; Mann--Whitney $p = 0.31$) - AI models consistently exhibited higher MAE on AI-authored essays than on human-authored ones across all blind and scheme conditions (e.g., committee blind: $+5.4$ marks). The AI models also assigned systematically lower raw scores to AI-authored text across nearly all conditions ($p < 0.05$), a differential not observed in the human marking. This is the exact opposite of what a self-leniency or stylistic-familiarity bias would predict~\cite{panickssery2024self_preference}. In the code dataset, AI models showed lower MAE on AI-generated code relative to student code, but AI-generated code also received genuinely lower human scores (mean $40.7$ vs.\ $64.3$ out of 70), making it difficult to disentangle marking accuracy from score-range compression. Taken together, the available evidence does not support the view that AI marking of AI-generated responses inflates validity estimates; if any authorship bias operates, it runs in the conservative direction.

\section{\label{sec:qual-errors}Qualitative analysis of marking errors}
This section addresses RQ2 by examining the mechanisms behind disagreements between human and AI markers. To better understand the specific mechanisms driving disagreements between human and AI markers, we qualitatively analysed the curriculum dataset, where blind, solution, and false solution conditions could be compared for the exact same items. Given that the exams lacked comparative conditions, the human essay baseline was too noisy for reliable error characterisation, and the scientific plots were already highly calibrated, our qualitative analysis focuses exclusively on the curriculum questions. We defined five distinct types of marking disagreement by isolating cases where all five AI models reached a unanimous verdict that contradicted the human reference mark:

\begin{enumerate}
\item \textbf{Blind over-crediting}: all models award full marks, human awards zero ($n = 13$).
\item \textbf{Blind under-crediting}: all models award zero, human awards full marks ($n = 9$).
\item \textbf{Over-crediting with solution}: all models award full marks despite a provided solution, human awards zero ($n = 2$).
\item \textbf{Under-crediting with solution}: all models award zero despite a provided solution, human awards full marks ($n = 12$).
\item \textbf{False solution anchoring}: all models award zero under the false solution condition for a question the human marked as fully correct ($n = 5$).
\end{enumerate}

Under blind conditions, 13 questions elicited unanimous full marks from the AI ensemble while the human reference awarded zero. Inspection of these cases reveals a consistent pattern. The GPT-3.5 responses typically contain numerically plausible answers with physically reasonable formatting, yet they are quantitatively incorrect - often due to an order-of-magnitude error, a missing unit conversion, or a misapplied physical assumption. Representative examples are shown in FIG.~\ref{fig:over-credit-examples}. Questions are lightly paraphrased from standard textbook sources due to copyright. For brevity, the full GPT-3.5 answers are not shown, but are available for all curriculum questions in the supplementary material. The copyrighted questions are not shared. These instances are exclusively short-answer numerical problems worth a single mark, where the GPT-3.5 response correctly adopts scientific notation and appropriate units. Rather than independently verifying the underlying calculation, all five models accepted the mathematically flawed answers at face value. This behaviour aligns with the broader finding that LLMs heavily weight surface-level structural plausibility when no explicit reference solution is available.

\begin{figure}[t]
\centering
\begin{tcolorbox}[colframe=black!25, colback=yellow!15, title={\small Blind over-crediting (all AI = full marks, human = 0)}]
{\small
\textbf{Q1.} Determine the total force on the bottom of a swimming pool ($28.0 \times 8.5\;\mathrm{m}$, depth $1.8\;\mathrm{m}$).\\
\textit{Solution:} $2.8 \times 10^7\;\mathrm{N}$\quad (1 mark)\\[4pt]
\textbf{Q2.} A man consumes 3000 kcal of food in one day. If he loses half this energy by evaporating water, calculate the mass of water evaporated.\\
\textit{Solution:} 2.59 kg\quad (1 mark)
}
\end{tcolorbox}
\caption{\label{fig:over-credit-examples}Representative paraphrased questions where all five AI models awarded full marks under blind conditions, but the human marker awarded zero. The GPT-3.5 responses contain plausible but quantitatively incorrect final values (e.g., calculating the gauge force of 4,198,320 N rather than the absolute force target of $2.8 \times 10^7\;\mathrm{N}$ for Q1, and 2.76 kg instead of 2.59 kg for Q2).}
\end{figure}

When the models were provided with the correct official solutions, instances of over-crediting dropped sharply from 13 to 2, a reduction of 85\%. Unlike the resolved cases, these two persistent errors highlight limitations in how LLMs utilize reference solutions (FIG.~\ref{fig:persistent-over}). The first involves the lenient crediting of flawed physics. For instance, in Q3, the AI awards full marks to a mathematically structured but physically incorrect derivation (calculating the extra force on the entire base of the jug rather than just the cork) even though the final numerical value ($\approx 5882\;\mathrm{N}$) fails to match the reference solution ($5.76 \times 10^{3}\;\mathrm{N}$). The second involves qualitative explanations. Here, the official solution provides a sample valid response but does not explicitly define the threshold for how rigorously the physics must be articulated. Consequently, the AI models leniently accept vague responses that the human markers did not credit.

\begin{figure}[t]
\centering
\begin{tcolorbox}[colframe=black!25, colback=yellow!15, title={\small Persistent over-crediting with solution (all AI = full marks, human = 0)}]
{\small
\textbf{Q3.} A host pours wine remnants into a jug and inserts a cork ($2.00\;\mathrm{cm}$ diameter) in direct contact with the wine. Find the force on the cork.\\
\textit{Solution:} $5.76 \times 10^{3}\;\mathrm{N}$\quad (1 mark)\\[4pt]
\textbf{Q4.} Ali uses a hot water bottle to keep warm and decides to switch to a heated seed pack. Suggest how a fair comparison could be made.\\
\textit{Solution:} Any one from: heat both in the same microwave for the same time at the same power\ldots\quad (1 mark)
}
\end{tcolorbox}
\caption{\label{fig:persistent-over}Representative paraphrased questions where all five AI models awarded full marks despite being provided with the correct solution, contradicting the human score of zero. In these cases, the models leniently accept flawed physical derivations despite mismatched final numbers (Q3) or reward vague qualitative phrasing (Q4), which the human markers did not credit.}
\end{figure}

Conversely, 9 questions produced unanimous AI zeros against a human verdict of full marks under blind conditions. Representative examples are shown in FIG.~\ref{fig:under-credit-examples}. These under-crediting errors cluster into three distinct categories. The first involves Fermi estimation, where questions require order-of-magnitude estimates (e.g., the number of books in a library or annual gasoline consumption). While the human marker accepts the GPT-3.5 answer as falling within a reasonable physical range, the models reject it, struggling to evaluate answers that lack a single, verifiable target. The second category comprises tolerance-dependent numerical answers. Here, GPT-3.5's answer differs from the textbook value by an amount within normal rounding or unit-conversion tolerances, but the models rigidly apply an exact-match criterion. Finally, under-crediting occurs in conceptual multi-step items requiring the explanation of a physical process. If GPT-3.5's phrasing is physically sound but structurally diverges from the model's internal formulation, the AI frequently penalises the response.

\begin{figure}[t]
\centering
\begin{tcolorbox}[colframe=black!25, colback=yellow!15, title={\small Under-crediting (all AI = 0, human = full marks)}]
{\small
\textbf{Q5.} A soccer field is 100.0\,m long; an American football field is 100.0\,yd. Which is longer, and by how much (in yards)?\\
\textit{Solution:} 9.4 yards\quad (1 mark)\\[4pt]
\textbf{Q6.} Estimate the number of books that can be shelved in a library with $3500\;\mathrm{m}^2$ of floor space (8 shelves high, 1.5\,m corridors, books 25\,cm tall and 4\,cm wide).\\
\textit{Solution:} $6 \times 10^5$ books\quad (1 mark)\\[4pt]
}
\end{tcolorbox}
\caption{\label{fig:under-credit-examples}Representative paraphrased questions where all five AI models awarded zero, while the human marker awarded full marks. These instances highlight model rigidity regarding rounding tolerances (e.g., rejecting an exact calculation of $9.361$ yards for Q5) and equivalent phrasings. Questions are lightly paraphrased from standard textbook sources.}
\end{figure}

Interestingly, whilst providing the correct solution reduced over-crediting errors it actually increased under-crediting errors, with the count rising from $n = 9$ to $n = 12$. This asymmetry - where solutions resolve 85\% of over-crediting but actually increase under-crediting - has a straightforward explanation. Over-crediting errors typically involve a model accepting a wrong answer because it looks structurally plausible; providing the correct answer gives the model a concrete numerical target that exposes the GPT-3.5 response's flaw. Under-crediting errors, however, involve a model rejecting a correct answer because it does not match a narrow, expected format. Providing the reference answer does not help in these cases because the GPT-3.5 response's physics is already correct. The model's failure stems not from an ignorance of the correct physics, but from an inability to recognise equivalent, approximately equal, or alternatively phrased responses. Indeed, providing the solution appears to introduce additional under-crediting cases, possibly because the explicit reference gives the model a stricter template against which to reject otherwise valid phrasings.

This asymmetry suggests that when a solution specifies an exact unambiguous target, providing it resolves most over-crediting disagreements. However, when the acceptable answer space is broad such as in Fermi estimates, rounding tolerance bands, or multiple valid phrasings of a physical concept the solution alone fails to supply the structural logic required to map diverse responses onto the marking criteria.

Finally, the strength of the anchoring effect was acutely demonstrated by five multi-mark questions that produced unanimous AI zeros under the false solution condition, despite the human awarding full marks. These items included multi-step calculations, derivations, and data interpretation questions. In each case, the false solution was generated via a minor mechanical perturbation of the correct answer (as detailed in Section~\ref{sec:methods}). All five models subsequently marked the perfectly correct GPT-3.5 response as wrong solely because it failed to match the false solution. While this represents a small subset of the broader false solution degradation reported in Section~\ref{sec:structured}, the unanimity across all five state-of-the-art models on multi-step derivations underscores a critical vulnerability that even when GPT-3.5's working is physically transparent and internally consistent, LLMs will reliably defer to a flawed reference solution rather than independently evaluating the physics.

\section{\label{sec:discussion}Discussion}
Overall, the results across the three task types present a coherent picture. For structured questions, models achieve moderate absolute accuracy and robust rank-order agreement under blind conditions, with both improving when a correct solution is supplied and degrading when a false solution is supplied. For essays, absolute accuracy is poor under blind conditions and improves superficially with anchoring, but rank-order agreement is indistinguishable from zero under all conditions ($\rho \approx 0.1$ blind, $\rho \approx 0$ scheme and anchored; all confidence intervals include zero). For scientific plots, both absolute accuracy and rank-order agreement are strong without any reference material. LLM marking is strongly task-dependent across the models tested, with the strongest agreement occurring where the assessment task provides observable features that both humans and models can map onto a relatively stable grading judgement.

\subsection{Task-dependent marking behaviour}
The pattern across the three task families is best read in terms of how explicitly each task defines a correct response. Structured questions are the clearest case. Even without an official mark scheme there is a well-defined target such as a correct numerical answer, a valid derivation, or a recognisable set of physically valid intermediate steps that can be checked against. This is why the models retain substantial rank-order agreement under blind marking and why supplying a correct solution improves absolute accuracy further. The observation is consistent with science-education research showing that assessment-construct features such as complexity, diversity, and structural transparency strongly influence automated-scoring accuracy~\cite{zhai2022construct_scoring}.

The code-plot task is not reducible to a single short answer but is similarly constrained in practice as a scientifically sound plot has identifiable properties such as sensible axes, labels, units, and scales, and consistency with the notebook context. Given the model judged a contextualised output embedded in a constrained workflow rather than an arbitrary graph in isolation the setting was structured enough for reliable relative judgements. Essays lie at the opposite end where the marker must balance content, relevance, argument structure, style, and coherence - judgements that are not illegitimate but are less transparent and less easily stabilised across raters. Once a task no longer supplies an explicit target, both human and AI marking become vulnerable to distributional heuristics and central-tendency effects. This is consistent with a strong task dependence in LLM marking behaviour across the models tested.

\subsection{Committee aggregation}
Across all three task types, the committee (mean of all individual models) provides modest but consistent benefits for absolute accuracy without substantially altering rank-order agreement. For structured questions and scientific plots, the committee tracks or marginally improves upon the best individual model (see Table~\ref{tab:discriminative-validity-questions} and Section~\ref{sec:plots}). The gains from aggregation are therefore incremental rather than transformative in settings where individual models already perform well, consistent with the general finding that ensemble averaging reduces idiosyncratic model errors but cannot compensate for systematic task-level limitations~\cite{verga2024poll}.

The essay results illustrate the limits of committee aggregation more starkly. Under the anchored condition, the committee achieves the lowest MAE (3.16) of any configuration, yet its rank-order agreement ($\rho = 0.034$) remains indistinguishable from zero. Averaging over models that individually lack rank-ordering ability does not recover discriminative signal. This suggests that committee aggregation is most useful as a variance reduction tool that is most effective when human marking is also consistent. We note that more sophisticated aggregation strategies - such as weighting by per-model reliability or using median rather than mean scoring - might yield different results, but given that the committee's primary limitation tracks the task-level validity ceiling rather than aggregation noise, such refinements are unlikely to alter the central conclusions.

\subsection{Interpretation of the essay results}
The essay findings should not be interpreted as evidence that LLMs are inherently poor at essay evaluation. A more precise interpretation is that this short-form holistic essay task provides a weak instrument for stable rank-ordering. Human markers exhibited almost no pairwise rank agreement ($\rho = 0.054$ on average; ICC2 $= 0.035$), indicating that the benchmark against which models are evaluated is already highly noisy at the single-rater level. In such conditions, reductions in MAE relative to the aggregated human score do not necessarily indicate that the model has learned to evaluate essay quality more validly; they may instead reflect alignment with the central tendency of the human score distribution. This behaviour is most clearly visible under the anchored condition. Providing exemplars compresses the AI score distribution and shifts its mean toward the aggregated human mean, but this occurs without any accompanying improvement in rank-order agreement, which remains indistinguishable from zero. In other words, the models better reproduce the distribution of human scores without improving their ability to distinguish between essays.

The human benchmark therefore forms part of the empirical result rather than serving solely as an external reference. For tasks involving substantial subjective judgement, the relevant question is not only whether AI scores agree with human scores, but also whether the human marking process itself supports meaningful ranking. If that condition is not met, apparent agreement with the human benchmark does not constitute evidence of valid assessment. This observation aligns with concerns raised in the automated essay scoring literature that distributional agreement with human benchmarks can be achieved without construct-valid discrimination~\cite{li2024aes_reflection,pack2024llm_aes_validity_reliability}.

\subsection{Implications for educators and assessment design}
These findings suggest a practical diagnostic for educators considering AI-assisted marking. Before asking whether LLM-as-a-judge can be used to mark a task, one should first establish whether human markers themselves assign stable and discriminative judgements for the task. If human raters cannot achieve acceptable rank agreement, an LLM is unlikely to recover valid ordering merely by being given more examples or guidance. In such cases, AI may still be useful for feedback, idea extraction, or preliminary moderation, but should not be treated as a dependable scoring instrument. AI-generated feedback on physics conceptual questions has been found to require only minor or no human modification in roughly 70\% of cases~\cite{wan2024feedback}, suggesting that even when AI scoring reliability is insufficient for summative grading, AI-assisted formative feedback may still be a defensible near-term application.

Conversely, where the task is tightly linked to identifiable criteria - as in structured questions and the Jupyter notebook-based plot task - AI marking appears much more defensible as an assistive tool. Even here, however, task complexity matters: error increases with available marks, implying that multi-step derivations remain harder to evaluate reliably than short, highly constrained responses. This points toward a layered deployment model, in which AI support is most appropriate for lower-mark tasks or tasks with clearly identifiable scoring criteria, with human oversight reserved for longer, more interpretive, or higher-stakes judgements.

A further practical implication concerns the use of exemplars. In human marking cultures, exemplar scripts are often employed to improve standardisation. The present results show that for LLM-as-a-judge exemplars can have qualitatively different effects depending on the task. In open-ended essay marking they may improve superficial agreement while undermining genuine discrimination. Educators should not assume that giving an LLM more marking support automatically improves validity. Polverini and Gregorčič provide practical guidance on how understanding LLM functioning can inform pedagogical deployment~\cite{polverini2024llms_physics_ed}, and Wulff reviews how AI can enhance language-related research and instruction in physics more broadly~\cite{wulff2024physics_language}.

\subsection{Reference dependence and authorship effects}
The clearest LLM-as-a-judge failure mode tested directly in this study is anchoring to supplied reference material. Providing a false solution degrades absolute accuracy for every model in the structured questions (Section~\ref{sec:structured}), and in the most extreme cases all five models award zero to a fully correct response solely because it diverges from the false reference (Section~\ref{sec:qual-errors}). The authorship analysis in Section~\ref{sec:self-preference} provides a separate check on whether AI-authored work is systematically favoured by AI markers. We do not observe such an effect: AI-authored work is marked no more leniently than human-authored work, and in this dataset is marked slightly more harshly. We do not directly test other LLM-judge biases such as verbosity or positional bias, and make no claims about them here.

\subsection{Governance and deployment}
These results reinforce the governance concerns raised in the Introduction. A system that produces low average error while failing to distinguish stronger from weaker work is not a valid assessor, even if its output appears well calibrated on aggregate. This is especially important in educational settings, where marks are not only summaries of performance but mechanisms for ranking, progression, and access to opportunity. Holmes et al.\ propose a widely cited ethical framework for AI in education covering fairness, accountability, and transparency~\cite{holmes2022ethics_aied}, and recent work on synthetic data generation in PER highlights both the promise of LLM-based methods and the methodological care required when AI systems interact with educational data~\cite{kieser2023data_augmentation}.

Recent discussions within the science education research community similarly emphasise that questions of validity, reliability, and ethical oversight must be addressed jointly when integrating AI into educational practice~\cite{kubsch2025workshop}. Within this framework, the present findings are more consistent with an assistive or auditing role for AI than with autonomous grading. LLMs may be useful for preliminary marking, second marking, anomaly detection, or feedback generation. In tasks where human inter-rater reliability is poor AI outputs should be treated as descriptive aids rather than authoritative judgements. The key governance question is therefore not ``does the AI look roughly right on average?'', but ``does this task admit a sufficiently valid and reliable marking process for AI outputs to be trusted in the first place?'' 

\begin{figure}[ht]
\centering
\includegraphics[width=\columnwidth]{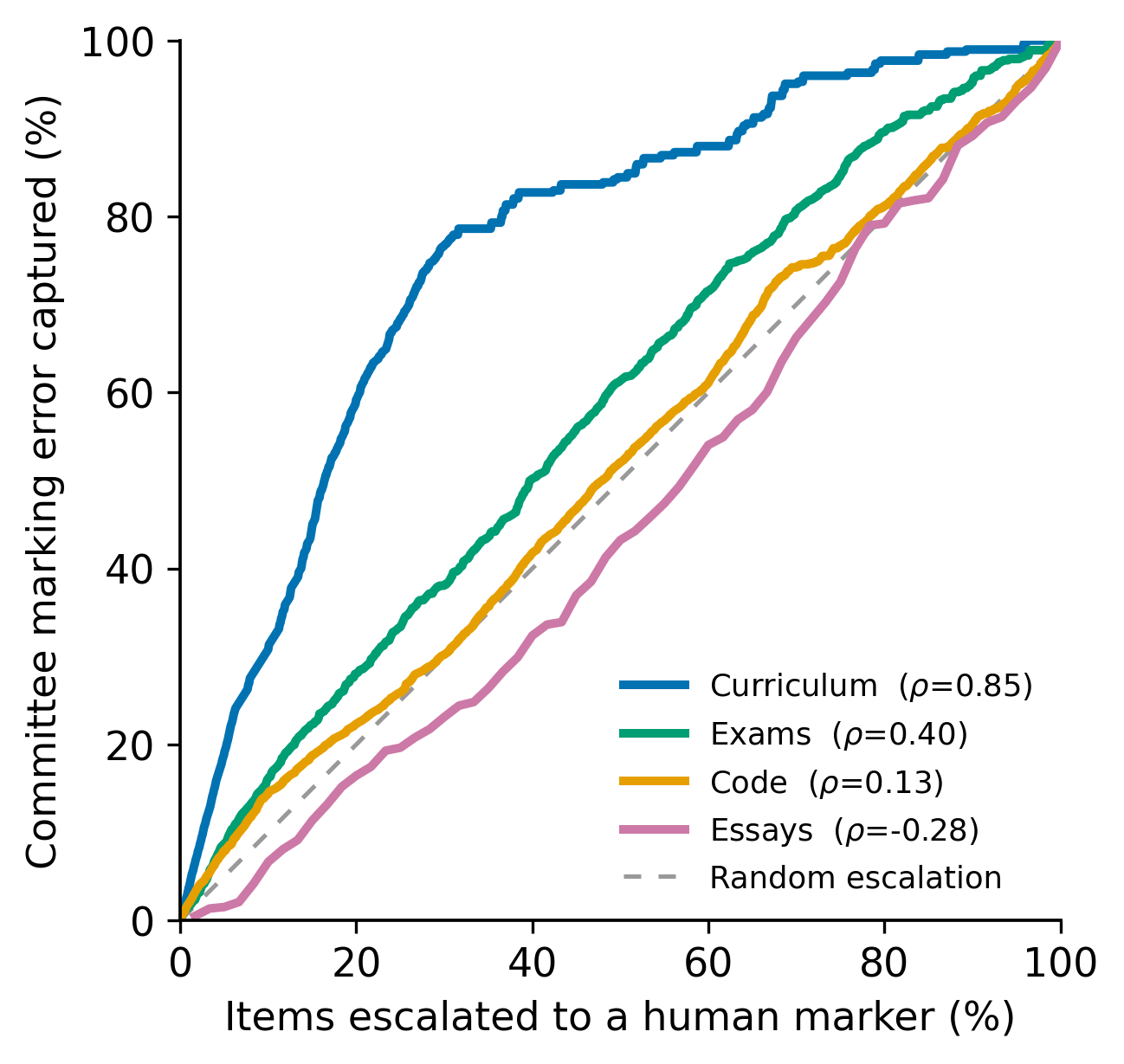}
\caption{\label{fig:escalation}Disagreement-triggered escalation. For each task family, items are ranked by committee disagreement (standard deviation of the individual model marks as a fraction of available marks) and escalated from most- to least-disagreed; the curve shows the fraction of total committee absolute error contained in the escalated items, assuming those items are then marked correctly by the human. The diagonal is the expectation under random
escalation, and $\rho$ is the Spearman correlation between committee disagreement and committee error. Curriculum and exam references are per question; the code reference is per sub-assignment (three image-capable models, human majority vote of three markers); essays use the aggregated human total. The exam reference is a single marker, which attenuates the achievable correlation.}
\end{figure}

A practical consequence for deployment is that committee disagreement can be tested as a triage signal for human review. FIG.~\ref{fig:escalation} asks whether the items on which models disagree most are also the items containing the largest committee marking errors. Because the committee mark is the mean of the individual model scores, this disagreement signal is available at no additional inference cost and can be used to elevate only the highest-disagreement items to a human marker. For the curriculum questions this is highly effective. Committee disagreement is strongly rank-correlated with committee error ($\rho = 0.85$, $n = 1151$), so escalating the 20\% of items with the widest model spread refers the items containing $59\%$ of all committee marking error to a human, and the fractional mean absolute error on the work still marked automatically falls from $0.127$ to $0.065$ (FIG.~\ref{fig:escalation}). One-fifth of the human effort therefore removes more than half of the residual error, and even a $5\%$ budget captures $19\%$. A department unwilling to release mark schemes could adopt blind committee marking as a first pass and reserve human attention for the high-disagreement tail.

This safeguard is not uniform across task types (FIG.~\ref{fig:escalation}). Its effectiveness tracks whether disagreement implies error (rather than how reliably a task can be marked) which holds only where an item has a unique verifiable target. When models disagree on a curriculum question scored against a single correct answer, at least one of them is simply wrong. As marking becomes graded or holistic the link weakens - exams $\rho = 0.40$, code $\rho = 0.13$, essays $\rho = -0.28$ - until disagreement carries essentially no information about error. The code plotting task is the clearest illustration as it is the most reliably marked task in this study (human-human $\rho \approx 0.8$ at the sub-item level, committee fractional MAE $0.16$ per individual plot), yet escalation by disagreement performs no better than random, because its residual errors are small band-level differences rather than the items on which the models happen to
diverge. For essays the curve falls below the random baseline, since with no stable reference model spread reflects diffuse central-tendency scoring rather than any recoverable error. Disagreement-triggered escalation is thus a cheap and effective human-in-the-loop mechanism for unique-target structured marking, but for graded or holistic tasks a fixed escalation budget gains little over random sampling, and human oversight must instead take the form of full marking or post-hoc moderation.

Finally, the practical deployment of LLM-as-a-judge marking requires consideration of inference costs, which vary considerably across providers. In the context of automated marking, a model typically generates only a few output tokens (a numerical score in a JSON object), meaning that the financial cost of the task is overwhelmingly driven by the input token volume. At the time of this study, API pricing for flagship models was roughly an order of magnitude higher per million input tokens than for alternative models, with input costs for Claude Opus 4.5 (\$5.00), GPT-5.2 (\$2.50), and Gemini 3 Pro (\$2.00) significantly outpacing those of DeepSeek-V3.2 (\$0.28) and Grok 4.1 (\$0.20). Given that the technical integration and inference speeds are broadly comparable across these platforms, institutions considering large-scale deployment must weigh these substantial cost differentials against performance and reliability, alongside relevant data security and privacy constraints.

\subsection{OCR and multimodal constraints}
While our evaluation provides a robust baseline for digital inputs, authentic physics assessments predominantly comprise handwritten derivations and spatial diagrams. Existing approaches - whether pipelined OCR-to-LaTeX or end-to-end vision-language models - show persistent performance gaps on handwritten mathematical input relative to digital text~\cite{kortemeyer2023toward, liu2024ai, nath2025fermat, baral2025drawedumath}, and the broader role of multimodal LLMs in science education remains under active development~\cite{bewersdorff2025mllm_science_ed}.

Our finding that LLMs achieve high rank-order agreement ($\rho > 0.83$) on scientific plots must therefore be interpreted alongside the broader literature on multimodal physics reasoning. Evaluating the structural and presentational quality of a generated plot on a fixed 0--5 descriptor scale - with access to the notebook context preceding the plot - is a relatively constrained task; direct semantic interpretation of complex visual physics data is not. Polverini and Gregorčič demonstrate that large multimodal models frequently fail to extract accurate information from kinematics graphs, and further evaluations on electromagnetism assessments reveal persistent difficulties with spatial reasoning~\cite{polverini2024evaluating, polverini2025performance}. This distinction between reliable contextual scale-based judgement and unreliable perceptual inference delineates where current models can and cannot yet be trusted. Until the visual encoders of multimodal models achieve greater perceptual robustness, deployment of AI grading for handwritten and visually complex physics submissions requires psychometrically calibrated, human-in-the-loop frameworks in which low-confidence interpretations automatically trigger human review~\cite{kortemeyer2024psychometrics}.

\subsection{\label{sec:limitations}Limitations and future work}
Several limitations should be noted. The structured-question datasets and the essay/plot datasets differ not only in task format but also in authorship composition and marking context, so the paper identifies robust cross-task contrasts without claiming that task format is the sole determinant of performance. The scientific plot results should be interpreted narrowly; they concern contextualised Jupyter notebook-based plot evaluation, not handwriting, arbitrary diagram interpretation, or general multimodal physics reasoning. And the essay dataset itself demonstrates that some assessment formats may be too psychometrically noisy to support strong claims about fine-grained ranking, whether by humans or by AI.

All main experiments were performed at temperature 0 to minimise stochastic variation and isolate systematic differences between tasks and prompting conditions. It remains possible that modest increases in temperature would alter score variance or anchoring behaviour for particular models, however preliminary work in Appendix~\ref{app-temp} suggests altering temperature does not significantly affect model behaviour. Further, given the scale of the differences observed between task types and prompt conditions, decoding settings are unlikely to overturn the central conclusion that LLM-as-a-judge marking is strongly task-dependent across the settings studied.

\section{Conclusion}
This study evaluated LLM-as-a-judge marking validity across three qualitatively different physics assessment formats: structured questions, short-form essays, and scientific plots. The results show that automated marking performance is strongly task-dependent. For structured questions and scale-guided plot evaluation, models achieve moderate absolute accuracy and strong rank-order agreement, often approaching the range of human–human disagreement. In contrast, essay marking exhibits fundamentally different behaviour: while anchored exemplars can improve distributional alignment and reduce mean absolute error, rank-order agreement remains indistinguishable from zero across all conditions, including under strong anchoring.

Across all three task types, LLM-as-a-judge validity is strongly task-dependent across the models tested. Where grading criteria can be explicitly linked to observable features of a response, LLMs can reproduce human-like ranking behaviour even under blinded conditions. Where assessment depends on holistic judgement and human raters themselves exhibit low reliability, agreement with human score distributions can be manufactured without meaningful discrimination between stronger and weaker work.

For educational practice, this implies that the key question is not simply whether LLM outputs appear reasonable on average, but whether the underlying assessment task admits a stable and interpretable marking process. 

In low-mark tasks with clearly identifiable scoring criteria, LLMs may plausibly support assistive roles such as preliminary marking, moderation, anomaly detection, or feedback generation. However, when human markers themselves fail to agree, automated scores should be interpreted cautiously and treated as descriptive aids rather than authoritative judgements. Future work should extend these findings to handwritten physics solutions and more complex multimodal assessments, where OCR errors and visual reasoning limitations may further constrain the reliability of automated marking systems.

\section{Acknowledgement}
This project received ethical approval from the Durham University physics ethics committee ref: \texttt{EDU-2023-03-14T14\_02\_18-hvxg44}.

\bibliographystyle{apsrev4-2}
\bibliography{references}

\appendix
\section{\label{app-code-plots}Examples of AI scientific plots}
\begin{figure}[ht]
\centering
\includegraphics[width=\columnwidth]{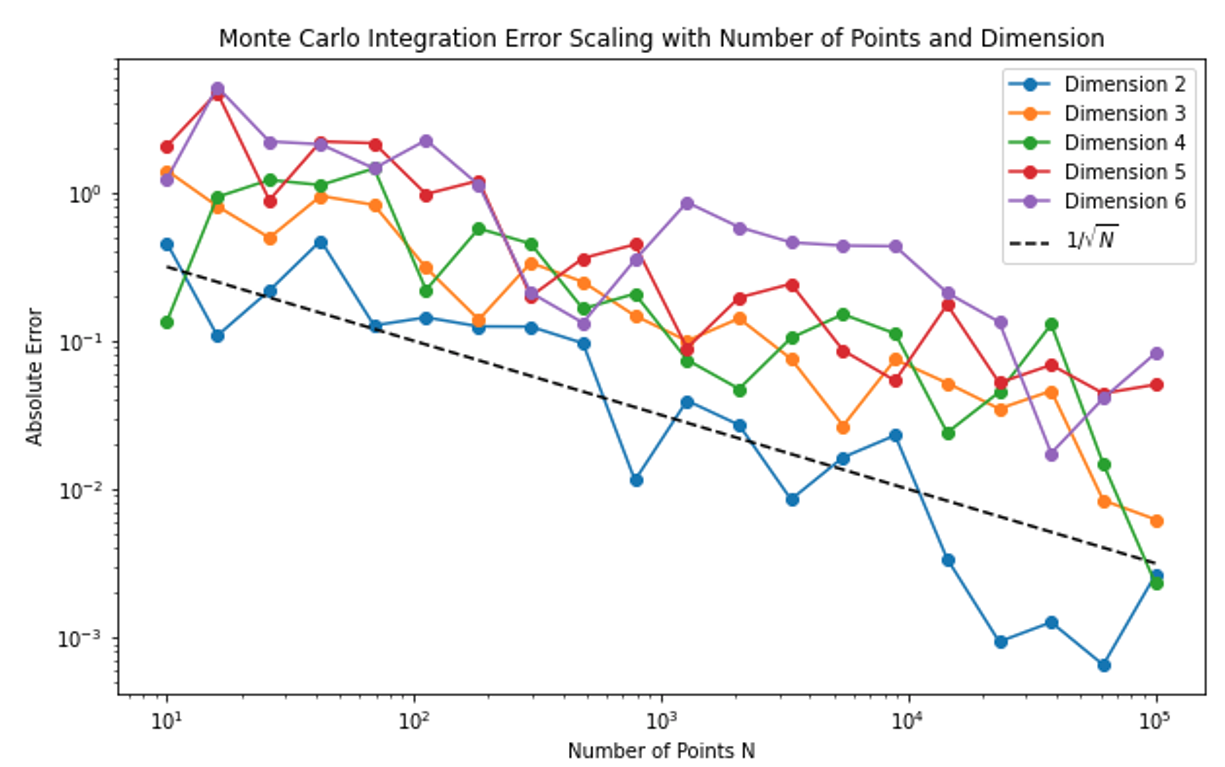}
\caption{\label{fig:gpt4}Plot produced by GPT-4 with prompt engineering, demonstrating the error scaling for Monte Carlo integration of an $n$-dimensional unit ball.}
\end{figure}

Examples of scientific plots produced by GPT-4 with prompt engineering, the highest average AI scoring group (FIG.~\ref{fig:gpt4}) and GPT-3.5 without prompt engineering, the lowest average AI scoring group (FIG.~\ref{fig:gpt35}) are shown below. These plots were produced for a laboratory task requiring the student to visually demonstrate the independence of integration error scaling from dimension versus the number of points used for Monte Carlo integration of an $n$-dimensional unit ball. Further details regarding the plotting tasks and generative methodologies are given in~\cite{coding-paper}.

\begin{figure}[ht]
\centering
\includegraphics[width=\columnwidth]{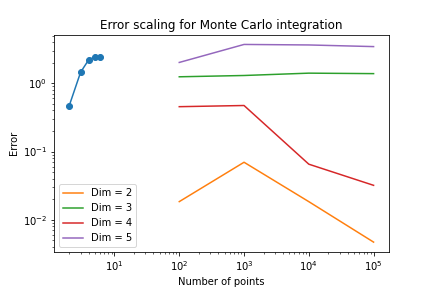}
\caption{\label{fig:gpt35}Plot produced by GPT-3.5 without prompt engineering, attempting to demonstrate the error scaling for Monte Carlo integration of an $n$-dimensional unit ball.}
\end{figure}

\section{\label{app-cot}Sensitivity to reasoning and Chain of Thought}
To assess whether the marking conclusions depend on the use of intermediate reasoning, we ran the DeepSeek-V3.2 marking pipeline under two additional configurations on the curriculum dataset ($n = 1151$) across all three structured-question conditions:

\begin{enumerate}
    \item \textbf{Explicit Chain of Thought.} DeepSeek-V3.2 was instructed to reason step by step before returning its JSON mark, at temperature $0$.
    \item \textbf{Native reasoner.} The DeepSeek-Reasoner model, which generates internal reasoning tokens prior to the final output, was given the same direct-output prompts as the baseline (no explicit Chain of Thought instruction). The reasoner model does not accept a temperature parameter.
\end{enumerate}

\begin{figure}[ht]
    \centering
    \includegraphics[width=\columnwidth]{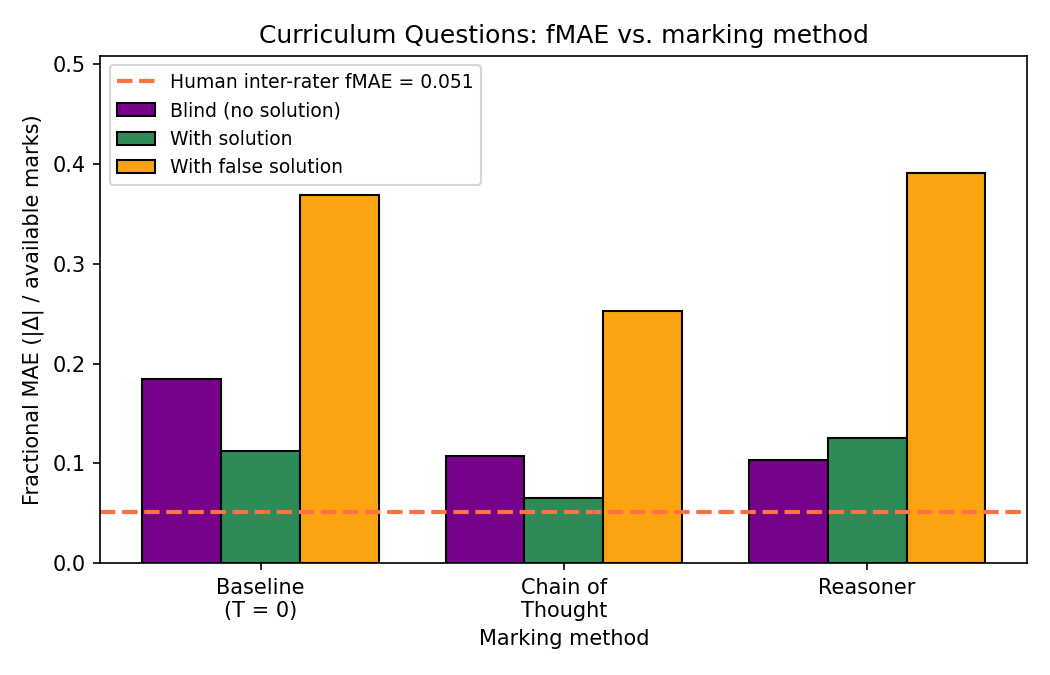}
    \caption{Fractional MAE for blind, solution-informed, and false solution marking conditions across three marking methods: baseline ($T = 0$), explicit Chain of Thought prompting, and the DeepSeek-Reasoner model. The dashed line shows human inter-rater fMAE ($0.051$) for reference.}
    \label{fig:cot_reasoner}
\end{figure}

As shown in FIG.~\ref{fig:cot_reasoner}, explicit Chain of Thought prompting reduced fMAE relative to the direct-output baseline across all three conditions (blind: $0.185 \to 0.108$; solution: $0.112 \to 0.066$; false solution: $0.369 \to 0.253$). The improvement is largest in the blind condition ($\approx 42\%$ reduction), consistent with the interpretation that step-by-step reasoning helps the model catch numerical errors it would otherwise overlook under direct-output prompting.

The DeepSeek-Reasoner model presents a contrasting pattern. Under blind marking it slightly outperforms the Chain of Thought result (fMAE $= 0.103$ vs.\ $0.108$), but under the solution condition it performs worse than the direct-output baseline ($0.125$ vs.\ $0.112$), and under the false solution condition it produces the highest fMAE of any configuration ($0.391$ vs.\ $0.369$ baseline). One possible interpretation is that the reasoner's extended internal reasoning causes it to engage more deeply with the provided reference material, amplifying rather than mitigating the anchoring effect. Where explicit Chain of Thought operates as a pre-output verification step that can surface discrepancies between the inputted answer-attempt and the model's own physics knowledge, the reasoner's native reasoning tokens may instead elaborate on the comparison to the supplied solution, reinforcing deference to the reference regardless of its correctness. This reading is consistent with the main-text finding that supplying a reference shifts the model from independent evaluation toward deference, and suggests that more reasoning does not by itself protect against that shift.

Critically, both configurations preserve the ordering of conditions established in the main text: providing a correct solution improves absolute accuracy relative to blind marking, and providing a false solution degrades it. The reasoner shows no clear overall benefit over the baseline. We therefore conclude that the central finding - that marking behaviour differs substantially by assessment task and by the quality of the supplied reference material - is robust to the prompting strategies tested here.

\section{\label{app-temp}Decoding temperature and marking reliability}
The main analysis uses temperature $T = 0$ throughout. Because hosted models are not guaranteed to be deterministic at $T = 0$ as batched inference and mixture-of-experts routing can introduce run-to-run variation we measured marking reliability directly rather than assuming it. We thus separate two questions: whether sampling temperature adds variance and whether marking is reproducible at fixed $T = 0$ across repeated runs using the curriculum and essay settings with the DeepSeek-V3.2 model.

\subsection{Sensitivity to sampling temperature}
\begin{figure}[ht]
\centering
\includegraphics[width=1\linewidth]{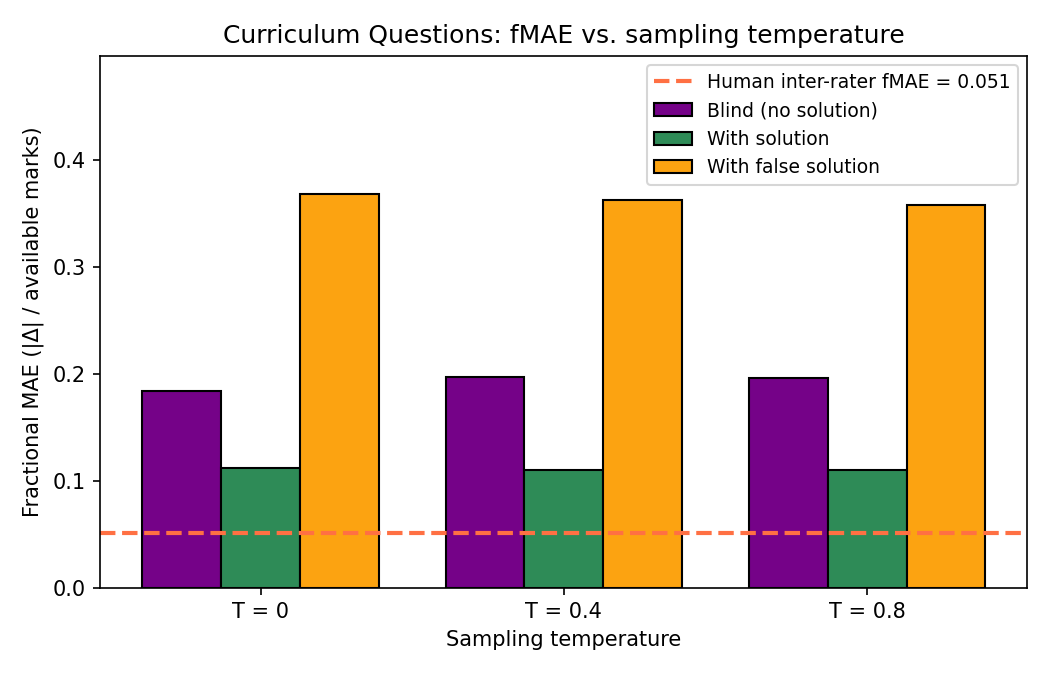}
\caption{Curriculum questions: fractional MAE for blind, solution-informed, and false-solution marking was stable across sampling temperatures $T = 0$, $0.4$, and $0.8$. Blind marking ranged from 0.185 to 0.197, solution-informed marking from 0.110 to 0.112, and false-solution marking from 0.358 to 0.369. The dashed line shows human inter-rater fMAE ($0.051$) for reference. Results are means over five runs for $T > 0$.}
\label{fig:temp_sensitivity}
\end{figure}

We ran the DeepSeek-V3.2 marking pipeline at $T = 0.4$ and $T = 0.8$ with five repeated runs each, comparing mean fMAE against the $T = 0$ baseline used in Section \ref{sec:structured} across all three structured-question conditions on the curriculum dataset ($n = 1151$). As shown in FIG.~\ref{fig:temp_sensitivity}, fMAE was stable across temperatures in all conditions: blind marking ranged from $0.185$ to $0.197$, solution-informed marking from $0.110$ to $0.112$, and false-solution marking from $0.358$ to $0.369$. Sampling temperature is therefore not a meaningful source of variance for structured-question marking.

\subsection{Reproducibility at fixed $T = 0$}
Temperature stability does not by itself establish that repeated $T = 0$ runs are identical as the exact decoding setup from a hosted model can change between providers. We therefore examined run-to-run agreement directly with five repeated $T = 0$ runs on the curriculum questions ($n = 1151$). Marking is near-deterministic but not perfectly so: $89.4\%$ (blind), $93.4\%$ (solution), and $89.5\%$ (false solution) of items received an identical mark on every one of the five runs. Crucially, the residual variation is small and structured rather than random: among non-identical items the mark differed by a single point in the large majority of cases (e.g.\ $100/123$ non-identical blind items), and the rare larger deviations concentrate on higher-mark, multi-step questions (items varying by $\geq 2$ marks carry $2.7$ available marks on average, versus $1.5$ across all items), where partial-credit allocation is itself a finer-grained judgement. The single-mark, multi-step locus of the residual variance is consistent with the five-run data at $T = 0.4$ and $T = 0.8$, where $91$--$97\%$ of items were likewise identical across runs (mean per-item standard deviation $0.015$--$0.052$ marks).

Essay marking behaves very differently. Repeating the $T = 0$ essay pipeline showed that script-level marks are \emph{not} reproducible: across six $T = 0$ runs, no script ($0/60$) received an identical total on every run, with a mean script-total standard deviation of $\approx 3$ marks out of $100$ (blind $3.4$, scheme $3.0$) and individual scripts varying by up to $15$ marks. At the level of individual essays ($/20$), only $\approx 10\%$ of marks were identical across runs. 

This run-to-run instability is intrinsic to the model rather than an artefact of the serving provider. When repeating the runs three times across two independent inference providers, the two providers ranked essays near-identically (Spearman $\rho = 0.93$) and differed in mean mark by only $\approx 2$ marks, so the $\approx 3$-mark within-script variation cannot be attributed to provider switching. This shows that DeepSeek-V3.2 is far more self-consistent than it is aligned with the human essay ranking. Across runs, the model reproduces its own essay ranking at $\rho \approx 0.85$ while agreeing with the human ranking at only $\rho \approx 0.1$ (Table~\ref{tab:essay_validity}). The instability therefore does not explain the absence of rank-order validity - the model reproduces a stable ordering that simply does not match the human one. Further, reproducibility itself tracks the structure of the task as marking is near-deterministic in the curriculum set ($89$--$94\%$ of marks identical across runs, with residual disagreements almost entirely of a single mark on multi-step items) and strongly non-deterministic in essays ($0\%$ of scripts identical, mean script-total standard deviation $\approx 3$ marks). The rank-order conclusions in the main text are unaffected, as essay $\rho$ remains near zero on every individual run (range $[-0.07, 0.26]$ across all six runs; Table~\ref{tab:essay_validity}); but the result shows that the assumption of $T = 0$ determinism, while safe for structured questions, does not hold for holistic marking.

\begin{table}[ht]
\centering
\caption{\label{tab:essay_validity}Essay marking reliability and validity across six repeated $T = 0$ runs (DeepSeek-V3.2, $n = 60$ scripts; runs 1--3 and 4--6 served by two independent inference providers). Rank-order agreement with the human reference (Spearman $\rho$) stays near zero on every run, while the model reproduces its own ranking across runs (self-consistency) far more strongly -- the marking is internally stable but shows weak agreement with the human ranking.}
\begin{ruledtabular}
\begin{tabular}{l c c c}
Condition
& \shortstack{$\rho$ vs.\ \\ human (range)}
& \shortstack{mean $\rho$ vs.\ \\ human}
& \shortstack{self-consistency\\ $\rho$} \\
\colrule
Blind   & $[0.07,\ 0.26]$   & $0.15$  & $0.82$ \\
Scheme  & $[-0.07,\ 0.06]$  & $-0.03$ & $0.85$ \\
\end{tabular}
\end{ruledtabular}
\end{table}

\section{\label{app-prompts}Prompts used}
\begin{figure*}[htp]
\centering
\begin{tcolorbox}[colframe=black!25, colback=blue!10]
\begin{verbatim}
SYSTEM:
You are an expert university physics examiner acting as an audit marker.
Assess scientific correctness.
Do not assume access to a mark scheme.
Do not invent missing context.
Follow the output format exactly.

USER:
QUESTION:
[QUESTION]

STUDENT RESPONSE:
[STUDENT RESPONSE]

AVAILABLE MARKS:
[AVAILABLE MARKS]

INSTRUCTIONS:
1. Assess the scientific correctness of the response.
2. Award a whole integer mark between 0 and the available marks
   (inclusive).
3. Do not assume access to a mark scheme.
4. Base your judgement on standard university physics expectations.

Return JSON:
{"awarded_marks": integer}
\end{verbatim}
\end{tcolorbox}
\caption{\label{fig:appendix-structured-blind}
Prompt template used for structured-question blind marking.}
\end{figure*}

\begin{figure*}[p]
\centering
\begin{tcolorbox}[colframe=black!25, colback=blue!10]
\begin{verbatim}
SYSTEM:
You are an expert university physics examiner acting as an audit marker.
Mark by comparing the student response to the provided model solution.
Do not invent missing context.
Be strict about what earns credit.
Follow the output format exactly.

USER:
QUESTION:
[QUESTION]

MODEL SOLUTION:
[MODEL SOLUTION]

STUDENT RESPONSE:
[STUDENT RESPONSE]

AVAILABLE MARKS:
[AVAILABLE MARKS]

INSTRUCTIONS:
1. Mark the student response by comparing it to the model solution.
2. Award a whole integer mark between 0 and [AVAILABLE MARKS] (inclusive).
3. Do not invent marking criteria beyond what is implied by the solution.
4. If the student gives an alternative correct method not shown in the solution, 
award credit if it is clearly correct.

Return JSON:
{"awarded_marks": integer}
\end{verbatim}
\end{tcolorbox}
\caption{\label{fig:appendix-structured-solution}
Prompt template used for structured-question marking with a supplied solution used for both the correct solution and the false solution cases.}
\end{figure*}

\begin{figure*}[p]
\centering
\begin{tcolorbox}[colframe=black!25, colback=blue!10]
\begin{verbatim}
SYSTEM:
You are an expert university physics examiner acting as an audit marker.
This is humanities-style physics / history / philosophy of science writing.
Assess accuracy, relevance, and argument quality.
Do not assume access to a mark scheme.
Do not invent missing context.
Be strict about what earns credit.
Follow the output format exactly.

USER:
PAPER: [PAPER CODE]

QUESTION ([QUESTION ID]):
[QUESTION TEXT]

STUDENT ESSAY RESPONSE:
[ESSAY TEXT]

AVAILABLE MARKS (for this essay only): 20

Return JSON exactly:
{"awarded_marks": integer}
\end{verbatim}
\end{tcolorbox}
\caption{\label{fig:appendix-essay-blind}
Prompt template used for blind essay marking.}
\end{figure*}

\begin{figure*}[p]
\centering
\begin{tcolorbox}[colframe=black!25, colback=blue!10]
\begin{verbatim}
SYSTEM:
You are an expert university physics examiner acting as an audit marker.
This is humanities-style physics / history / philosophy of science writing.
Mark by comparing the student response to the provided mark scheme /
model answer.
Do not invent missing context.
Be strict about what earns credit.
Follow the output format exactly.

USER:
PAPER: [PAPER CODE]

QUESTION ([QUESTION ID]):
[QUESTION TEXT]

MARK SCHEME / MODEL ANSWER (guidance):
[MODEL ANSWER OR GUIDANCE TEXT]

STUDENT ESSAY RESPONSE:
[ESSAY TEXT]

AVAILABLE MARKS (for this essay only): 20

Return JSON exactly:
{"awarded_marks": integer}
\end{verbatim}
\end{tcolorbox}
\caption{\label{fig:appendix-essay-scheme}
Prompt template used for essay marking with the scheme/guidance text.}
\end{figure*}

\begin{figure*}[p]
\centering
\begin{tcolorbox}[colframe=black!25, colback=blue!10]
\begin{verbatim}
CALIBRATION EXAMPLES (for score calibration only):
These are real student responses with known human mean marks (out of 20).
Use them ONLY to calibrate what ~low / ~mid / ~high marks look like.
Do NOT copy phrases from them. Do NOT mention them.

--- $\approx$5th percentile (human mean 12.2/20) ---
Human mean mark (out of 20): 12.2
Student response:
[EXEMPLAR RESPONSE 1]

--- $\approx$25th percentile (human mean 13.0/20) ---
Human mean mark (out of 20): 13.0
Student response:
[EXEMPLAR RESPONSE 2]

--- $\approx$50th percentile (human mean 13.4/20) ---
Human mean mark (out of 20): 13.4
Student response:
[EXEMPLAR RESPONSE 3]

--- $\approx$75th percentile (human mean 13.6/20) ---
Human mean mark (out of 20): 13.6
Student response:
[EXEMPLAR RESPONSE 4]

--- $\approx$95th percentile (human mean 14.2/20) ---
Human mean mark (out of 20): 14.2
Student response:
[EXEMPLAR RESPONSE 5]

END CALIBRATION EXAMPLES.

NOW MARK THIS NEW RESPONSE:

PAPER: [PAPER CODE]

QUESTION ([QUESTION ID]):
[QUESTION TEXT]

MARK SCHEME / MODEL ANSWER (guidance):
[MODEL ANSWER OR GUIDANCE TEXT]

STUDENT ESSAY RESPONSE:
[ESSAY TEXT]

AVAILABLE MARKS (for this essay only): 20

Return JSON ONLY:
{"awarded_marks": integer}
\end{verbatim}
\end{tcolorbox}
\caption{\label{fig:appendix-essay-anchor}
Prompt template used for essay marking with scheme plus anchored calibration exemplars.}
\end{figure*}

\begin{figure*}[p]
\centering
\begin{tcolorbox}[colframe=black!25, colback=blue!10]
\begin{verbatim}
SYSTEM:
You are an expert university physics examiner marking plotting outputs.
You do NOT have a mark scheme.
Mark each plot on correctness (scientific correctness + whether it
communicates clearly).
Return ONLY valid JSON.

USER:
ASSIGNMENT/TASK TEXT (what the student was asked to do):
[ASSIGNMENT TEXT]

INSTRUCTIONS:
Use this 0--5 anchor scale:
5 = scientifically correct and clear; axes/units/labels/scale sensible;
    presentation clean.
4 = scientifically correct; minor presentation issues (small
    label/legend/scale problems).
3 = mostly correct but at least one substantive weakness affecting
    interpretation.
2 = major scientific/presentation problems; plot gives wrong/unclear
    takeaway.
1 = mostly wrong or key requirements missing.
0 = blank/irrelevant/unreadable.

Return JSON:
{"awarded_marks": integer}

[IMAGE INPUT: submitted plot]
\end{verbatim}
\end{tcolorbox}
\caption{\label{fig:appendix-code-prompt}
Prompt template used for marking of scientific plots on the 0–5 descriptor scale.}
\end{figure*}

\end{document}